%% file: pratik.tex
\documentclass{article}

\input preamble.tex

\def\sscoin{%
  \leavevmode
  \vtop{\offinterlineskip %\bfseries
    \setbox0=\hbox{\scriptsize S}%
    \setbox2=\hbox to\wd0{\hfil\hskip-.03em
    \vrule height .3ex width .08ex\hskip .08em
    \vrule height .3ex width .08ex\hfil}
    \vbox{\copy2\box0}\box2}}
\newcommand\affil[2]{%
  \begingroup
  \renewcommand\thefootnote{}\footnote{\llap{$\hbox{}^{#1}\hbox{}$}#2}%
  \addtocounter{footnote}{-1}%
  \endgroup
}
\newcommand\markonly[1]{%
$\hbox{}^{\mbox{\kern4.5pt,\kern0.75pt #1}}$
}

\advance\oddsidemargin by -0.525in
\advance\textwidth by 1.05in
\advance\topmargin by -0.625in
\advance\textheight by 1.25in

\title{\vspace{-0.5in}An Empirical Analysis of Image-Based Learning 
Techniques for Malware Classification}

\author{Pratikkumar Prajapati\footnote{pratikkumar.prajapati@sjsu.edu}\ \ \ \ 
Mark Stamp\footnote{mark.stamp@sjsu.edu}\markonly{\sscoin}
}
\date{}

\begin{document}

\maketitle

\vglue-0.35in

\affil{\sscoin}{Department 
of Computer Science,
San Jose State University,
San Jose, California}

\abstract
In this paper, we consider malware classification
using deep learning techniques and image-based features. 
We employ a wide variety of 
deep learning techniques, including multilayer perceptrons (MLP), 
convolutional neural networks (CNN), long short-term memory (LSTM),
and gated recurrent units (GRU). Amongst our CNN experiments,
transfer learning plays a prominent role---specifically, we test the VGG-19
and ResNet152 models. As compared to previous work, 
the results presented in this paper are based on a larger and more
diverse malware dataset, we consider a wider array of features,
and we experiment with a much greater variety of learning techniques. 
Consequently, our results 
are the most comprehensive and complete that
have yet been published.

\section{Introduction}\label{sect:intro}

Traditionally, malware detection and classification has relied on 
pattern matching against signatures extracted from specific malware samples. 
While simple and efficient, signature scanning is easily defeated by a number of 
well-known evasive strategies. 
This fact has given rise to statistical and machine learning based techniques, 
which are more robust to code modification. In response, malware writers 
have developed advanced forms of malware that alter statistical and 
structural properties of their code, which  
can cause statistical models to fail.

In this paper, we compare deep learning (DL) 
models for malware classification.
For most of
our deep learning models, we use image-based features,
but we also experiment with
opcode features. The 
DL models consider include a wide variety of 
neural networking techniques, including multilayer perceptrons (MLP),
several variants of convolutional neural networks (CNN), and 
vanilla recurrent neural networks (RNN),
as well as the advanced RNN architectures known as
long short-term memory (LSTM) and gated recurrent units (GRU).
We also experiment with a complex stacked model that combines
both LSTM and GRU. In addition, we consider transfer learning, 
in the form of the ResNet152 and VGG-19 architectures.

The remainder of this paper is organized as follows. In Section~\ref{sect:back}
we provide relevant background information, including a discussion of related work,
an overview of the various learning techniques considered, and we introduce
the dataset used in this research. Section~\ref{sect:exp} is the heart of the paper,
with detailed results from a wide variety of malware classification
experiments. Section~\ref{sect:con} concludes the paper and 
provides possible directions for future work.

\section{Background}\label{sect:back}

In this section, we discuss related work and we introduce the various
learning techniques that are considered in this research. We also
discuss the dataset that we use in our malware classification experiments.
In addition, we provide the specifications of the hardware and software
that we use to conduct the extensive set of experiments that are summarized in 
Section~\ref{sect:exp}.

\subsection{Related Work}

To the best of our knowledge, image-based analysis was 
first applied to the malware problem in~\cite{PB1}, where
high-level ``gist'' descriptors are used as features. % to successfully classify malware.
More recently,~\cite{PB2} confirmed the results in~\cite{PB1}
and presented an alternative deep learning approach that produces equally 
good---if not slightly better---results,
without the extra work required to extract gist descriptors.
%which is significant when scoring samples.

%A direct comparison to classic machine learning techniques 
%seems to be lacking in previous work, making it difficult to determine the 
%comparative advantages and disadvantages of deep learning image-based 
%analysis in the malware domain. One of the goals of this paper is to
%provide such a comparison.

Transfer learning, where the output layer of an existing pre-trained DL model
is retrained for a specific task, is often used in image analysis. Such an approach
allows for efficient training, as a new model can take advantage of a vast
amount of learning that is embedded in the pre-trained model.
Leveraging the power of transfer learning has been shown to yield 
strong image-based malware detection and classification results~\cite{PB2}.

There is a vast malware analysis literature involving classic
machine learning techniques. Representative examples 
include~\cite{AustinFJS13},
\cite{BaysaLS13},
\cite{DamodaranTVAS17},
\cite{SinghTVAS16},
%\cite{LeeAS15},
\cite{TodericiS13},
\cite{WongS06}.
Intuitively, we might expect models based on image analysis to be 
somewhat stronger and more robust,
as compared to models that rely on opcodes, byte $n$-grams,
or similar statistical features that are commonly used
in malware research.

The work presented in this paper can be considered an extension of the
work in~\cite{BPTS}, where image-based transfer learning is applied
to the malware classification problem. We have extended this previous
work in multiple dimensions, including a larger, more challenging,
and more realistic dataset. In addition, we perform much more
experimentation with a much wider variety of techniques, and we 
consider a large range of hyperparameters in each case.

\subsection{Learning Techniques}

In this section, we provide a brief introduction to each of the learning
techniques considered in this paper. Additional details on most of the 
learning techniques discussed here can be found 
in~\cite{Stamp_DL_survey}, which includes examples of  
relevant applications of the techniques. We provide additional references
for the techniques discussed below that are not considered 
in~\cite{Stamp_DL_survey}.

\subsubsection{Multilayer Perceptron}

A perceptron computes a weighted sum of its components
in the form of a hyperplane, and
based on a threshold, a perceptron can be used to define a classifier. 
%Consequently, a perceptron defines a hyperplane.
It follows that a perceptron cannot provide ideal separation in cases
where the data itself is not linearly separable. This is a severe
limitation, as something as elementary as the XOR function
is not linearly separable.

A multilayer perceptron (MLP) is an artificial neural network that includes
multiple (hidden) layers in the form of perceptrons. 
Unlike a single layer perceptron, MLPs are not 
restricted to linear decision boundaries, and hence an MLP can accurately
model more complex functions. The relationship between
perceptrons and MLPs is very much analogous to the relationship
between linear support vector machines (SVM) and SVMs based
on nonlinear kernel functions.
 
Training an MLP would appear to be challenging
since we have hidden layers between the
input and output, and it is not clear how changes to the
weights in these hidden layers will affect each other or the output.
Today, MLPs are generally trained using backpropagation. 
The discovery that backpropagation can be used for training neural networks
was a major breakthrough that made deep learning practical.

%\begin{figure}[!htb]
%\centering
%    \input figures/figMLP.tex
%\caption{MLP with two hidden layers}\label{fig:mini_MLP}
%\end{figure}

\subsubsection{Convolutional Neural Network}

Generically, artificial neural networks use fully connected layers.
The advantage of a fully connected layer is that it
can deal effectively with correlations 
between any points within training vectors. 
However, for large training vectors, fully connected layers
are infeasible, due to the vast number of weights that must be learned.

In contrast, a convolutional neural network (CNN), 
is designed to deal with local structure. A convolutional layer
cannot be expected to perform well when significant information is not local. 
The benefit of CNNs is that convolutional layers can be 
trained much more efficiently than fully connected layers,
due to the reduced number of weights.

For images, most of the important structure (edges and gradients, for example) 
is local. Hence, CNNs are an ideal tool for
image analysis and, in fact, CNNs were developed precisely 
for image classification.
However, CNNs have performed well in a variety of other problem domains.
In general, any problem for which local structure predominates is a candidate for CNNs.
%In addition to images, local structure is crucial in fields such as
%text analysis and speech analysis, among many others.

\subsubsection{Recurrent Neural Network}

MLPs and CNNs are feedforward neural networks, that is,
the data feeds directly through the network,
with no ``memory'' of previous feature vectors. In a feedforward network,
each input vector is treated independently of other input vectors. 
While feedforward networks are appropriate for many problems,
they are not well-suited for dealing with sequential data.

In some cases, it is necessary for a classifier to have memory.
Suppose that we want to tag parts of speech
in English text (i.e., noun verb, etc.), this is not feasible
if we only look at words in isolation. For example, the word ``all'' can be 
an adjective, adverb, noun, or pronoun, and this can only
be determined by considering its context. 
A recurrent neural network (RNN) provides
a way to add memory (or context) to a feedforward neural network.

RNNs are trained using a variant of backpropagation known as 
backpropagation through time (BPTT). A problem that is particularly acute
in BPTT is that the gradient calculation tends to be become unstable,
resulting in ``vanishing'' or ``exploding'' gradients. To overcome these
problems, we can limit the number of time steps, but this also
serves to limit the utility of RNNs. 
Alternatively, we can use specialized
RNN architectures that enable the gradient to flow over long
time periods. Both long short-tem memory and 
gated recurrent units 
are examples of such specialized RNN architectures.
We discuss these two RNN architectures next.

\subsubsection{Long Short-Term Memory}

Long short-term memory (LSTM) networks are a class of RNN 
architectures that are designed to deal with long-range dependencies.
That is, LSTM can deal with extended ``gaps'' between the appearance 
of a feature and the point at which it is needed by the model. In plain
vanilla RNNs this is generally not possible, due
vanishing gradients.

The key difference between an LSTM and a
generic vanilla RNN is that an LSTM includes an additional
path for information flow. That is, in addition to the the hidden state,
there is a so-called cell state that can be used to, in effect, store information
from previous steps. The cell state is designed to
serve as a gradient ``highway'' during backpropagation. In this way,
the gradient can ``flow'' much further back with less chance that it will 
vanish (or explode) along the way.

As an aside, we
note that the LSTM architecture has been one of the most 
commercially successful learning techniques ever developed. 
Among many other applications, LSTMs play a critical role
in Google Allo~\cite{allo}, Google Translate~\cite{translate}, 
Apple's Siri~\cite{iBrain}, and Amazon Alexa~\cite{alexa}. 

\subsubsection{Gated Recurrent Unit}

Due to its wide success, many variants on the LSTM architecture
have been considered.
Most such variants are slight, with only minor changes
from a standard LSTM.  However, a gated recurrent unit (GRU) 
is a fairly radical departure from an LSTM.
Although the internal state of a GRU is somewhat complex 
and less intuitive than that of an LSTM, 
there are fewer parameters in a GRU. As a result, it is
easier to train a GRU than an LSTM, 
and consequently less training data is required. 

\subsubsection{ResNet152}

Whereas LSTM uses a complex gating structure to ease gradient flow,
a residual network (ResNet) defines additional connections that correspond 
to identity layers. 
These identity layers allow a ResNet model to, in effect,
skip over layers during training, which serves 
to effectively reduce the depth when training
and thereby mitigate gradient pathologies.
Intuitively, ResNet is able to train deeper networks by 
training over a considerably shallower network in the initial
stages, with later stages of training serving to flesh out the intermediate connections. 
This approach was inspired by pyramidal cells in the brain, which
have a similar characteristic, in the sense that they bridge 
``layers'' of neurons~\cite{pyramid}.

ResNet152 is a specific deep ResNet architecture that has been pre-trained 
on a vast image dataset. As one of our two examples of transfer learning,
we use this architecture, which includes
an astounding~152 layers. That is, we use the ResNet152
model, where we only retrain the output layer specifically 
for our malware classification problem.

\subsubsection{VGG-19}

VGG-19 is a 19-layer convolutional neural network that has been pre-trained
on a dataset containing more than~$10^6$ images~\cite{vgg19}. This architecture
has performed well in many contests, 
and it has been generalized to a variety of image-based problems.
Here, we use the VGG-19 architecture and pre-trained model as one
of our two examples of transfer learning for image-based malware classification.

\subsection{Dataset}

Our dataset consists of~20 malware families.
Three of these malware families, namely,
Winwebsec, Zeroaccess, and Zbot,
are from the Malicia dataset~\cite{nappa2015malicia}, 
while the remaining~17 families are taken from the massive 
malware dataset discussed in~\cite{bigdata}. 
This latter dataset is almost half a 
terabyte and contains more than~500,000 malware samples in the form of 
labeled executable files. 

Table~\ref{tab:families} lists the~20 families used in this research, along with
the type of malware present in each family.
Next, we briefly discuss each of these~20 malware families.

\begin{table}[!htb]
\begin{center}
\caption{Type of each malware family}\label{tab:families}
{\footnotesize
\begin{tabular}{cc|cc}\midrule\midrule
\textbf{Family} & \textbf{Type} & \textbf{Family}  & \textbf{Type} \\
\midrule
Adload~\cite{adload}      & Trojan Downloader            & Obfuscator~\cite{obfuscator}  & VirTool             \\
Agent~\cite{agent}       & Trojan        & Onlinegames~\cite{Onlinegames} & Password Stealer            \\
Alureon~\cite{alureon}     & Trojan           & Rbot~\cite{rbot}        & Backdoor            \\
BHO~\cite{bho}         & Trojan          & Renos~\cite{renos}    & Trojan Downloader            \\
CeeInject~\cite{CeeInject}   & VirTool            & Startpage~\cite{startpage}   & Trojan            \\
Cycbot~\cite{cycbot}    & Backdoor            & Vobfus~\cite{vobfus}      & Worm            \\
DelfInject~\cite{DelfInject}  & VirTool            & Vundo~\cite{vundo}       & Trojan Downloader            \\
FakeRean~\cite{FakeRean}    & Rogue            & Winwebsec~\cite{winwebsec}   & Rogue            \\
Hotbar~\cite{hotbar}      & Adware            & Zbot~\cite{zbot}        & Password Stealer            \\
Lolyda~\cite{lolyda}   & Password Stealer            & Zegost~\cite{zegost}  & Backdoor     \\
\midrule\midrule 
\end{tabular}
}
\end{center}
\end{table}

\begin{description}

\item[\textbf{Adload}] downloads an executable file, stores it remotely, 
executes the file, and disables proxy settings~\cite{adload}. 

\item[\textbf{Agent}]
downloads trojans or other software from a remote server~\cite{agent}. 

\item[\textbf{Alureon}]
exfiltrates usernames, passwords, credit card information, and 
other confidential data from an infected system~\cite{alureon}. 

\item[\textbf{BHO}]
can perform a variety of actions, guided by an attacker~\cite{bho}. 

\item[\textbf{CeeInject}]
uses advanced obfuscation to avoid being detected by antivirus software~\cite{CeeInject}. 

\item[\textbf{Cycbot.G}]
connects to a remote server, exploits vulnerabilities, and spreads through 
a backdoor~\cite{cycbot}. 

\item[\textbf{DelfInject}]
sends usernames, passwords, and other personal 
and private information to an attacker~\cite{DelfInject}. 

\item[\textbf{FakeRean}]
pretends to scan the system, notifies the user of supposed issues, 
and asks the user to pay to clean the system~\cite{FakeRean}. 

\item[\textbf{Hotbar}]
is adware that shows ads on webpages and installs additional adware~\cite{hotbar}. 

\item[\textbf{Lolyda}]
sends information from an infected system and monitors the system. 
It can share user credentials and network activity with an attacker~\cite{lolyda}. 

\item[\textbf{Obfuscator}]
tries to obfuscate or hide itself to defeat malware detectors~\cite{obfuscator}.

\item[\textbf{Onlinegames}]
steals login information and tracks user 
keystroke activity~\cite{Onlinegames}. 

\item[\textbf{Rbot}]
gives control to attackers via a backdoor that can be used to access information or
launch attacks, and it serves as a gateway to infect additional sites~\cite{rbot}.

\item[\textbf{Renos}]
downloads software that claims the system has spyware and asks for a payment to 
remove the nonexistent spyware~\cite{renos}. 

\item[\textbf{Startpage}]
changes the default browser homepage and can perform other
malicious activities~\cite{startpage}. 

\item[\textbf{Vobfus}]
is a worm that downloads malware and spreads through USB drives or other 
removable drives~\cite{vobfus}. 

\item[\textbf{Vundo}]
displays pop-up ads and it can download files. It uses advanced techniques to 
defeat detection~\cite{vundo}.

\item[\textbf{Winwebsec}]
displays alerts that ask the user for money to 
fix nonexistent security issues~\cite{winwebsec}.

\item[\textbf{Zbot}]
is installed through email and shares a user's personal information with attackers.
In addition, Zbot can disable a firewall~\cite{zbot}.

\item[\textbf{Zegost}]
creates a backdoor on an infected machine~\cite{zegost}.

\end{description}

The number of samples per malware family for the various features
are given in Table~\ref{tab:samps}. The ``Binaries'' lists the number
of binary executable files available, the ``Images'' column lists the
number of binaries that were successfully converted to images, and
the ``Opcodes'' column lists the number of samples from which a sufficient number
of opcodes were extracted.
From the table we see that~$26{,}413$ samples are used in
our image-based experiments and~$25{,}901$ samples are used
in our opcode-based experiments. 

\begin{table}[!htb]
\begin{center}
\caption{Samples per malware family}\label{tab:samps}
{\footnotesize
\begin{tabular}{c|ccc}\midrule\midrule
\multirow{2}{*}{\textbf{Family}} & \multicolumn{3}{c}{\textbf{Samples}} \\
 & \textbf{Binaries} & \textbf{Images} & \textbf{Opcodes} \\ \midrule
Adload & 1050 & 1050 & 1044 \\
Agent & \zz842 & \zz842 & \zz817 \\
Alureon & 1328 & 1328 & 1327 \\ 
BHO & 1176 & 1176 & 1159 \\
CeeInject & \zz894 & \zz894 & \zz886 \\
Cycbot & 1029 & 1029 & 1029 \\
DelfInject & 1146 & 1146 & 1097 \\
Fakerean & 1063 & 1063 & 1063 \\
Hotbar & 1491 & 1491 & 1476 \\
Lolyda & \zz915 & \zz915 & \zz915 \\
Obfuscator & 1445 & 1445 & 1331 \\
Onlinegames & 1293 & 1293 & 1284 \\
Rbot & 1017 & 1017 & 817 \\
Renos & 1312 & 1312 & 1309 \\
Startpage & 1136 & 1136 & 1084 \\
Vobfus & \zz926 & \zz926 & \zz924 \\
Vundo & 1793 & 1793 & 1784 \\
Winwebsec & 3651 & 3651 & 3651 \\
Zbot & 1786 & 1786 & 1785 \\
Zeroaccess & 1120 & 1120 & 1119 \\
\midrule
Total & $26{,}413$ & $26{,}413$ & $25{,}901$ \\
\midrule\midrule 
\end{tabular}
}
\end{center}
\end{table}

\subsection{Hardware}

Table~\ref{tab:machine} lists the hardware configuration of
the machine used for the experiments reported in this paper.
This machine was assembled for the purpose of training deep
learning models and it is highly optimized for this task.

\begin{table}[!htb]
\caption{Hardware characteristics}\label{tab:machine}
\centering
{\small
\begin{tabular}{c|l|l}\midrule\midrule
\textbf{Feature} & \textbf{Description} & \textbf{Details} \\ \midrule
\multirow{4}{*}{CPU} 
& Brand and model & Intel i9-9940X \\
& Clock Frequency & 3.30GHz \\
& Number of threads & 28 \\
& Cache & 19.25 MB Intel Smart Cache \\
	\midrule
\multirow{4}{*}{CPU liquid cooling}
& Brand and model & Corsair Hydro Series H115i PRO RGB \\
& Fan Speed & 1200 RPM \\
& Fan size & 140mm \\
& Radiator size & 280mm \\
	\midrule
\multirow{3}{*}{DRAM}
& Brand and model & Corsair CMK32GX4M2A2666C16 \\
& Speed & 2666MHz \\
& Capacity & $16 \times 8 = 128\mbox{GB}$ \\
	\midrule
Motherboard 
& Brand and model & ASUS WS x299 Sage \\
	\midrule
\multirow{7}{*}{GPU}
& Brand and model & Nvidia Titan RTX \\
& Total Video Memory & 24 GB GDDR6 \\
& Tensor Cores & 576 \\
& CUDA Cores & 4608 \\
& Base Clock (MHz) & 1350 MHz \\
& Single-Precision Performance & 16.3 TFLOPS \\
& Tensor Performance & 130 TFLOPS \\
	\midrule
\multirow{3}{*}{Storage}
& Brand and model & Sabrent 2TB Rocket NVMe \\
& Read Speed & 3400 MB/s \\
& Write Speed & 2750 MB/s \\
\midrule\midrule
\end{tabular}
}
\end{table}

\subsection{Software}

For our deep learning neural network experiments,
we have used PyTorch~\cite{torch}. In addition, for general data processing 
and related operations, we employ both Numpy~\cite{numpy} 
and Pandas~\cite{pandas}. In addition, all code that was developed as part
of this project is available at~\cite{github}.

\section{Deep Learning Experiments and Results}\label{sect:exp}

In this section, we present results of a wide variety of neural network based
experiments. 
First, we consider MLP experiments, followed by CNN experiments, and then
RNN experiments. We consider a large number of CNN and RNN cases.
We conclude this section with a pair of models based on
transfer learning. The MLP, CNN, and transfer learning models
are based on image features, while the RNN experiments use 
opcode sequences.

We consider
various different sizes for images, in each case using square images.
To generate a square image from an executable, we first specify a width~$N$, with the height
determined by the size of the sample. We then resize the image so that it is~$N\times N$,
which has the effect of stretching or shrinking the height, as required. 

\subsection{Multilayer Perceptron Experiments}

We experimented with various perceptron-based neural networks.
The model we present here uses square input image
and has four hidden layers, each using 
the popular rectified linear unit (relu) activation function. The output from the 
final hidden layer is passed to a fully connected output layer. The output layer
is used to classify the sample---since we have~20 classes of malware 
in our dataset, the output vector is 20-dimensional.
The hyperparameters used for these MLP experiments are
given in Table~\ref{tab:classMLP}.

\begin{table}[!htb]
\caption{MLP model parameters}\label{tab:classMLP}
\centering
{\small
	\begin{tabular}{c|c|c|c|c}\midrule\midrule
		\multirow{2}{*}{\textbf{Classifier}} 
			& \multirow{2}{*}{\textbf{Hyperparameter}} 
				& \multirow{2}{*}{\textbf{Tested values}} 
				& \multicolumn{2}{c}{\textbf{Accuracy}} \\
			& & & \textbf{Train} & \textbf{Test} \\  
				\midrule
		%%%%% Don't see any difference between these models
%		\multirow{4}{*}{MLP\un model1} 
%			& \texttt{image\un dim} & $[64,\textbf{128}]$ & \multirow{4}{*} {$0.9604$} 
%				&  \multirow{4}{*} {$0.8579$} \\
%			& \texttt{learning\un rate} & $[0.001,\textbf{0.0001}]$ & \\
%			& \texttt{batch\un size} & $256$ & \\
%			& \texttt{epochs} & $50$ & \\
%				\midrule
		\multirow{4}{*}{MLP} 
			& \texttt{image\un dim} & $[64,\textbf{128}]$ 
				& \multirow{4}{*} {$0.9529$} &  \multirow{4}{*} {$0.8644$} \\
			& \texttt{learning\un rate} & $[0.001, \textbf{0.0001}]$ & \\
			& \texttt{batch\un size} & $256$ & \\
			& \texttt{epochs} & $50$ & \\
				\midrule\midrule
	\end{tabular}
}
\end{table}

\begin{figure}[!htb]
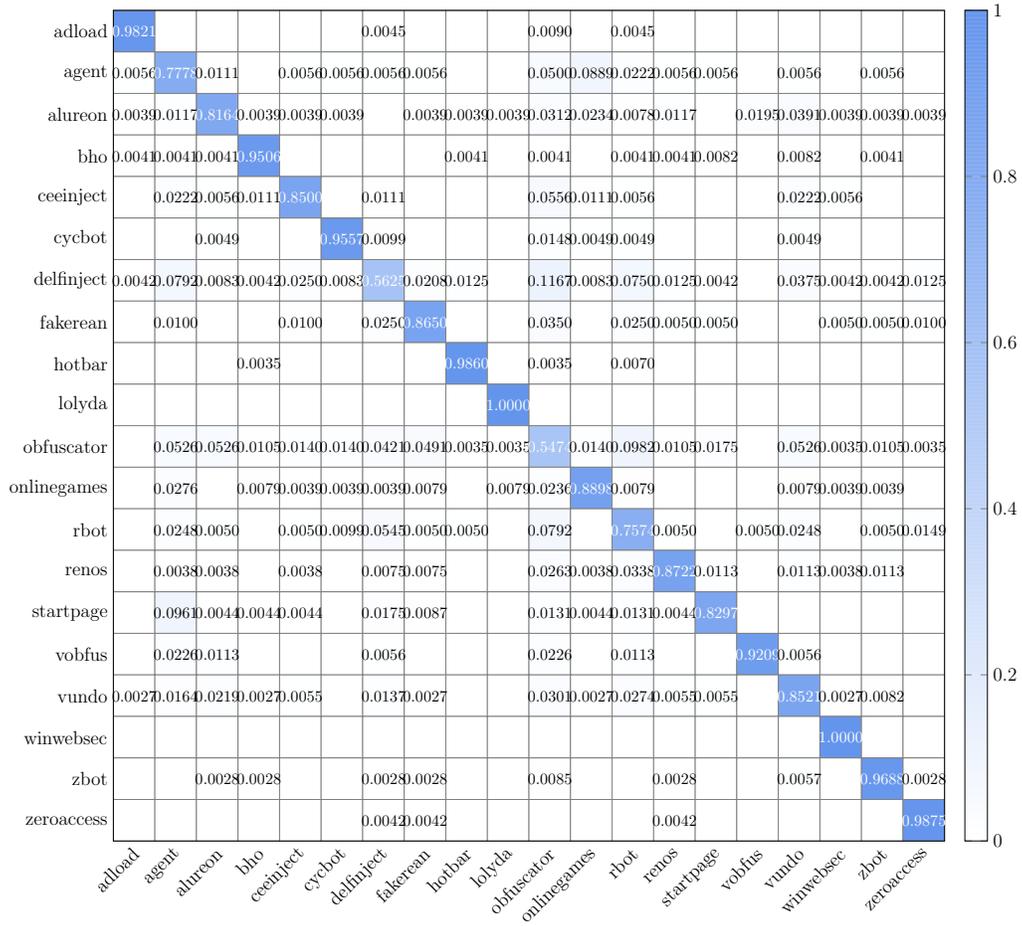

	\centering
	\input figures/conf_ANNMalware_Model2.tex
	\caption{Confusion matrix for MLP experiment}
	\label{fig:conf_mlp}
\end{figure}

Figure~\ref{fig:conf_mlp} gives the confusion matrix for the best results
obtained in our MLP experiments. 
The hyperparameters used for this best case are those 
shown in boldface in Table~\ref{tab:classMLP}. In this case,
the DelfInject and Obfuscator families have the lowest detection
rates, with both only slightly above~50\%\ accuracy. The overall 
accuracy is~0.8644.

\subsection{Convolutional Neural Network Experiments}

We have conducted a large number of convolutional neural network (CNN)
experiments. In this section we first discuss CNN experiments
based on 2-dimensional images. Then we consider 1-dimensional CNN
experiments, where the malware images are vectorized. We also present results
%%%%% How many opcodes ?????
for CNN experiments using opcodes extracted from PE files,
as opposed to forming images based on the raw byte values
in the executable files. The opcodes were extracted using objdump,
and we use the resulting mnemonic opcode sequence (eliminating
operands, labels, etc.) as features.
The hyperparameters tested for
all of these CNN experiments are given in Table~\ref{tab:classCNN}.

\begin{table}[!htb]
\caption{CNN model parameters}\label{tab:classCNN}
\centering
{\small
	\begin{tabular}{c|c|c|c|c}\midrule\midrule
		\multirow{2}{*}{\textbf{Classifier}} 
			& \multirow{2}{*}{\textbf{Hyperparameter}} 
				& \multirow{2}{*}{\textbf{Tested values}} 
				& \multicolumn{2}{c}{\textbf{Accuracy}} \\
			& & & \textbf{Train} & \textbf{Test} \\  
				\midrule
		\multirow{4}{*}{CNN 2-d} % \multirow{4}{*}{CNN 2-d small} 
			& \texttt{image\un dim} & $[64, 128, 256, \textbf{1024}]$ 
				& \multirow{4}{*} {$0.9294$} &  \multirow{4}{*} {$0.8955$} \\ 
			& \texttt{learning\un rate} & $[\textbf{0.001}, 0.0001]$ & \\
			& \texttt{batch\un size} & $256$ & \\
			& \texttt{epochs} & $50$ & \\
			%Pratik best performing model was with epochs = 8
				\midrule
		%%%%% Parameters here just a subset of model 1
%		\multirow{4}{*}{CNN\un model2} 
%			& \texttt{image\un dim} & $[64, \textbf{128}]$ & \multirow{4}{*} {$0.9918$} 
%				&  \multirow{4}{*} {$0.8804$} \\ 
%			& \texttt{learning\un rate} & $[0.001, \textbf{0.0001}]$ & \\
%			& \texttt{batch\un size} & $256$ & \\
%			& \texttt{epochs} & $50$ & \\
%				\midrule
		\multirow{4}{*}{CNN 1-d} % \multirow{4}{*}{CNN 2-d large}
			& \texttt{image\un dim} & $[\textbf{1024}, 2048, 4096, 8192]$ 
				& \multirow{4}{*} {$0.8445$} &  \multirow{4}{*} {$0.8664$} \\ 
			& \texttt{learning\un rate} & $[\textbf{0.001}, 0.0001]$ & \\
			& \texttt{batch\un size} & $256$ & \\
			& \texttt{epochs} & $20$ & \\
				\midrule
		\multirow{10}{*}{CNN 1-d refined} % \multirow{10}{*}{CNN 1-d} 
			& \texttt{conv1d\un 1\un out\un channel} & $[64, \textbf{128}]$ 
				& \multirow{10}{*} {$0.8538$} &  \multirow{10}{*} {$0.8932$} \\ 
			& \texttt{conv1d\un 1\un kernel\un size} & $[\textbf{16}, 32]$ & \\
			& \texttt{conv1d\un 1\un stride} & $[2,\textbf{8}]$ & \\
			& \texttt{conv1d\un 2\un out\un channel} & $[\textbf{32}, 64,128]$ & \\
			& \texttt{conv1d\un 2\un kernel\un size} & $[\textbf{8},16]$ & \\
			& \texttt{conv1d\un 2\un stride} & $[\textbf{2}, 4]$ & \\
			& \texttt{image\un dim} & $4096$ & \\
			& \texttt{learning\un rate} & $0.001$ & \\
			& \texttt{batch\un size} & $512$ & \\
			& \texttt{epochs} & $15$ & \\
				\midrule
		\multirow{7}{*}{CNN opcode} 
			& \texttt{opcode\un length} & $[500, \textbf{5000}]$ 
				& \multirow{7}{*} {$0.8418$} &  \multirow{7}{*} {$0.8282$} \\ 
			& \texttt{num\un filters} & $[3,6,\textbf{9}]$ & \\
			& \texttt{filter\un size} & $[[12, 6], [6, 12], \textbf{[12, 24]}]$ & \\
			& \texttt{embedding\un dim} & $[128, \textbf{512}]$ & \\
			& \texttt{learning\un rate} & $0.001$ & \\
			& \texttt{batch\un size} & $256$ & \\
			& \texttt{epochs} & $50$ & \\
				\midrule\midrule
	\end{tabular}
}
\end{table}

\subsubsection{2-Dimensional Image CNNs}

Based on 2-dimensional image features, we 
test the CNN model hyperparameters listed under ``CNN 2-d'' in
Table~\ref{tab:classCNN}. 
All of these 2-d CNN experiments use
two convolutional layers and three fully connected layers. 
The first convolutional layer takes as input a square gray-scale image
with one channel and outputs data with~12 channels using 
a kernel size of three, padding of two, and a stride of one. 
A relu activation and max pooling is applied to the result
before passing it to the second convolutional layer. This second layer 
outputs data with~16 channels, with the other parameters being the
same as the first convolutional layer. Again, relu activation and max pooling 
is applied before passing data to the first fully connected layer. 
This first fully connected layer outputs a vector of dimension~120. 
After applying relu activation, the data is passed to the second fully 
connected layer, which reduces the output to a~90 dimensional vector.
Finally, relu activation is again applied and the data passes to the last 
fully connected layer, which is used to classify the sample, and hence 
is~20-dimensional. For all image sizes less than~1024, 
we execute our CNN 2-d models for~50 epochs; 
for~$1024\times 1024$ images, we use~8 epochs due to the 
costliness of training on these large images.

%CNNMalware\un Model2 is similar to CNNMalware\un Model1 with following changes. 
%The first convolution layer outputs 15 channels using a kernel size of 15. The first 
%fully connected layer produces a vector of dimension which is 80\% of input image 
%dimension. The second fully connected layer produces a vector of dimension 
%which is 40\% of input image dimension. And this vector is passed to the last layer 
%for classification.

The best overall accuracy obtained for our CNN 2-d experiments is~0.8955. 
Figure~\ref{fig:conf_CNN_small} gives the confusion
matrix for the best case. We note that the Obfuscator family is again
the most difficult to distinguish. 

\subsubsection{Vectorized Image CNNs}

Recent work has shown promising results for malware classification
using 1-dimensional CNNs on ``image'' data~\cite{Mugdha}.
Consequently, we experiment with flattened images, that is,
we use images that are one pixel in height. A possible advantage of this
approach is that 2-dimensional results can depend on the width chosen for
the images. We perform two sets of such experiments, which we
denote as CNN 1-d and CNN 1-d refined, the latter of which considers
additional fine-tuning parameters. The hyperparameters tested
for this two case are given in Table~\ref{tab:classCNN}.

Our CNN 1-d model uses two 1-dimensional convolutional layers,
followed by three fully connected layers. The first convolution layer 
takes in an image with one channel and outputs data with~28 channels 
based on a kernel size of three. The second convolutional layer outputs 
data with~16 channels and again uses a kernel of size three. 
The first fully connected layer
outputs a vector of~120 dimensions, which is reduced to~90 dimensions 
by the second fully connected which, in turn, is reduced to~20
dimensions by the third (and last) fully connected layer. 
We have applied relu activations in all layers.

The confusion matrix for our best CNN 1-d case
is give in Figure~\ref{fig:conf_CNN_1d}. The overall
accuracy in in this case is~0.8664. A handful of families (Agnet, Alureon, DelfInject,
Obfuscator, and Rbot) have accuracies below~80\%, which represents the 
majority of the loss of accuracy.

The CNN 1-d refined tests use the same basic setup as our CNN 1-d experiments,
but includes different selections of hyperparameters. As expected,
these additional parameters improved on the CNN 1-d case,
as the best overall accuracy attained for our CNN 1-d refined 
experiments is~0.8932. Qualitatively, the CNN 1-d refined results
are similar (per family) to the CNN 1-d experiments, 
%%%%% ?????? Seems odd to omit the confusion matrix for the better case ??????
so we have omitted the confusion matrix for this case.

\subsubsection{Opcode Based CNNs}

We also apply 2-d CNNs to opcode features. For each malware sample, we use the
first~$N$ opcodes from each binary file, where $N\in\{500,5000\}$. 
We also experiment with various other parameters, as indicated in
Table~\ref{tab:classCNN}.

\begin{figure}[!htb]
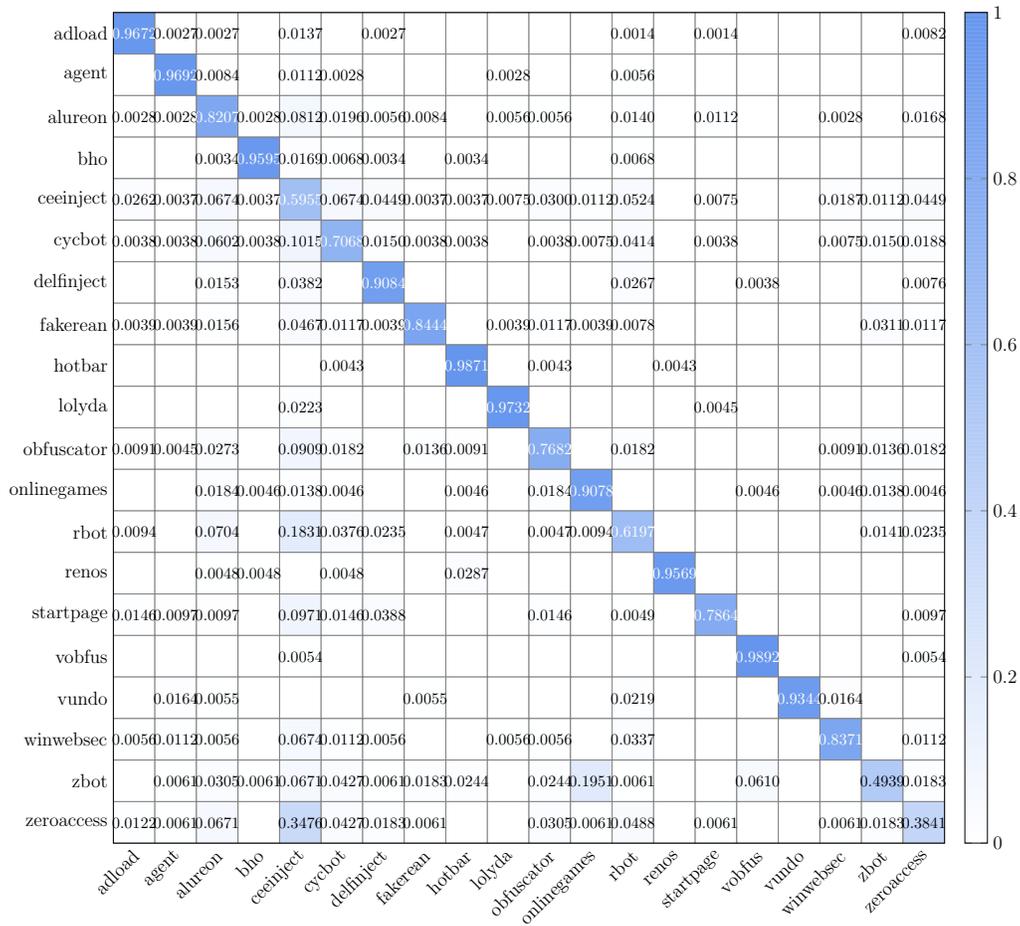

	\centering
	\input figures/conf_CNNMalware_Model5.tex
	\caption{Confusion matrix for opcode-based CNN experiment}
	\label{fig:conf_CNN_opcode}
\end{figure}

The results for the best choice of parameters for our opcode-based CNN 
experiments are summarized in the confusion matrix in 
Figure~\ref{fig:conf_CNN_opcode}. Perhaps not surprisingly, the results
in this case are relatively weak, with an overall accuracy of~0.8282.
However, it is interesting to note from the confusion matrix that some of the
families that are consistently misclassified at high rates by 
image-based CNN models are classified with high accuracy by
this opcode-based approach. For example, DelfInject is classified at
no better than about~71\%\ in our previous CNN experimetns, but
it is classified with greater than~90\%\ accuracy using the opcode-based
features.
%%%%% ????? How about an SVM to combine outputs from all CNN models 
%%%%% or even to combine results from more models ?????

\subsection{Recurrent Neural Networks}

Next, we consider a variety of experiments based on 
various recurrent neural network (RNN) architectures.
Specifically, we employ plain vanilla RNN, LSTM, and GRU models.
We also consider a complex LSTM-GRU stacked model.
The hyperparameters tested in these experiments are
summarized in Table~\ref{tab:classRNN}.

\begin{table}[!htb]
\caption{RNN model parameters}\label{tab:classRNN}
\centering
{\small
	\begin{tabular}{c|c|c|c|c}\midrule\midrule
		\multirow{2}{*}{\textbf{Classifier}} 
			& \multirow{2}{*}{\textbf{Hyperparameter}} & \multirow{2}{*}{\textbf{Tested values}} 
				& \multicolumn{2}{c}{\textbf{Accuracy}} \\
			& & & \textbf{Train} & \textbf{Test} \\  
				\midrule
		\multirow{7}{*}{RNN} 
			& \texttt{embedding\un dim} & $[\textbf{256}, 1024]$ 
				& \multirow{7}{*} {$0.7710$} &  \multirow{7}{*} {$0.7294$} \\ 
			& \texttt{hidden\un dim} & $[\textbf{256}, 1024]$ & \\
			& \texttt{num\un layers} & $[\textbf{1},3]$ & \\
			& \texttt{directional} & $\texttt{[\textbf{uni-dir}, bi-dir]}$ & \\
			& \texttt{learning\un rate} & $0.001$ & \\
			& \texttt{batch\un size} & $128$ & \\
			& \texttt{epochs} & $20$ & \\
				\midrule
		\multirow{7}{*}{LSTM} 
			& \texttt{embedding\un dim} & $[\textbf{256}, 1024]$ 
				& \multirow{7}{*} {$0.9362$} &  \multirow{7}{*} {$0.8916$} \\ 
			& \texttt{hidden\un dim} & $[\textbf{256}, 1024]$ & \\
			& \texttt{num\un layers} & $[\textbf{1},3]$ & \\
			& \texttt{directional} & $\texttt{[\textbf{uni-dir}, bi-dir]}$ & \\
			& \texttt{learning\un rate} & $0.001$ & \\
			& \texttt{batch\un size} & $128$ & \\
			& \texttt{epochs} & $20$ & \\
				\midrule
		\multirow{7}{*}{GRU} 
			& \texttt{embedding\un dim} & $[\textbf{256}, 1024]$ 
				& \multirow{7}{*} {$0.9411$} &  \multirow{7}{*} {$0.9003$} \\ 
			& \texttt{hidden\un dim} & $[\textbf{256}, 1024]$ & \\
			& \texttt{num\un layers} & $[\textbf{1},3]$ & \\
			& \texttt{directional} & $\texttt{[\textbf{uni-dir}, bi-dir]}$ & \\
			& \texttt{learning\un rate} & $0.001$ & \\
			& \texttt{batch\un size} & $128$ & \\
			& \texttt{epochs} & $20$ & \\
				\midrule
		\multirow{7}{*}{Stacked} 
			& \texttt{embedding\un dim} & $[256, \textbf{1024}]$ 
				& \multirow{7}{*} {$0.9525$} &  \multirow{7}{*} {$0.8990$} \\ 
			& \texttt{hidden\un dim} & $[256, \textbf{1024}]$ & \\
			& \texttt{num\un layers} & $[\textbf{1},3]$ & \\
			& \texttt{directional} & $\texttt{[\textbf{uni-dir}, bi-dir]}$ & \\
			& \texttt{LG} & $\texttt{[\textbf{True}, False]}$ & \\
			& \texttt{learning\un rate} & $0.001$ & \\
			& \texttt{batch\un size} & $128$ & \\
			& \texttt{epochs} & $20$ & \\
				\midrule\midrule
	\end{tabular}
}
\end{table}

\subsubsection{Vanilla RNN, LSTM, and GRU}

We have trained our plain vanilla RNNs, LSTMs, and GRU-based models
using~20 epochs in each case, with a learning rate of~0.001, 
a batch size of~128, and based on the first~500 opcodes 
from each malware sample. We performed multiple experiments with various 
other parameters, as given in Table~\ref{tab:classRNN}.
In addition, we have applied a dropout layer with~0.3 probability 
for all models with more than one layer.
%%%%% ????? Why no confusion matrices for vanilla RNN and LSTM ?????

\begin{figure}[!htb]
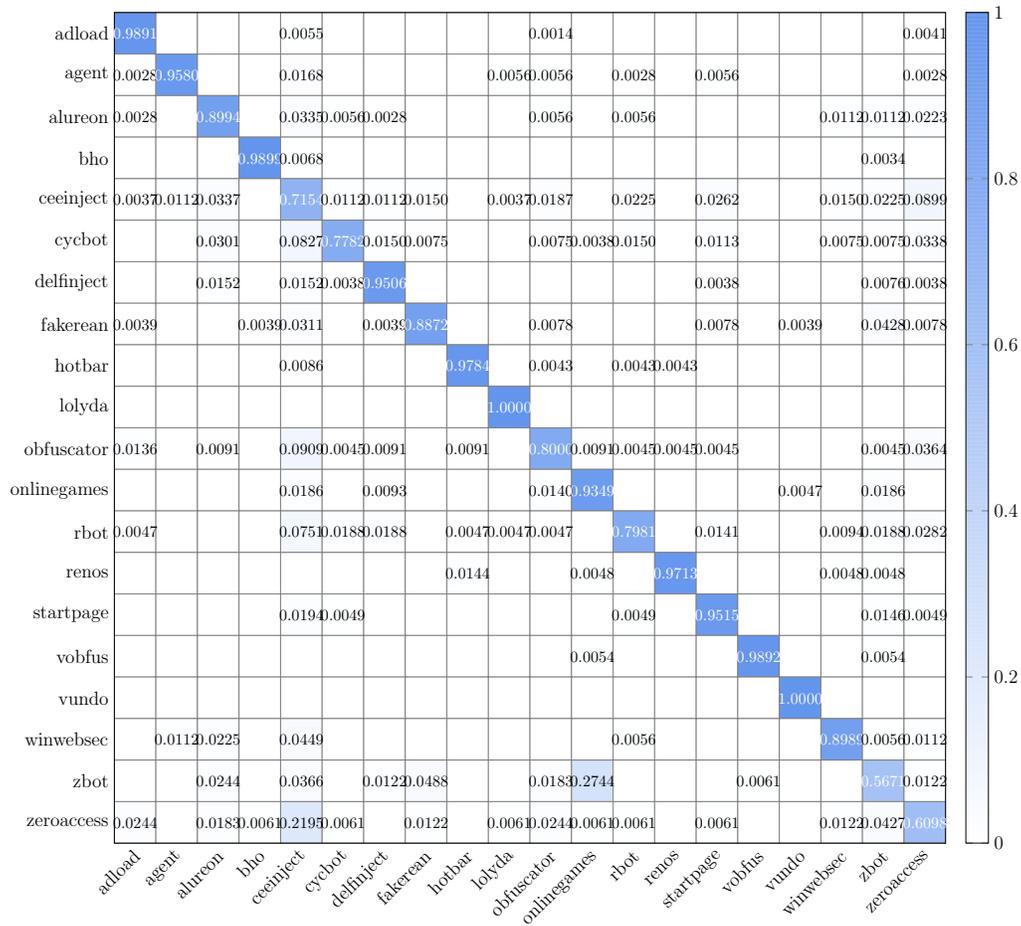

	\centering
	\input figures/conf_GRUMalware_Model1.tex
	\caption{Confusion matrix for GRU experiment}
	\label{fig:conf_GRU}
\end{figure}

The vanilla RNN experiments performed poorly, with an overall accuracy
of just~0.7294, and hence we omit the confusion matrix for this case.
On the other hand, both the LSTM and GRU models perform well,
with accuracies of~0.8916 and~0.9003, respectively. 
The confusion matrix for the GRU case is given in 
Figure~\ref{fig:conf_GRU}. Since the LSTM results are
so similar, we omit the LSTM confusion matrix.
From Figure~\ref{fig:conf_GRU}, we see that, qualitatively, 
the results of our GRU experiments
more closely match those of the CNN opcode-based experiments
than the CNN image-based experiments. However, quantitatively,
our GRU opcode-based experiments yield significantly 
better results than our CNN opcode-based experiments.

\subsubsection{Stacked LSTM-GRU Model}

As in~\cite{prat_5}, we have also experimented with 
stacked LSTM and GRU layers. The experiments
in this paper test more parameters and we use a 
larger dataset, as compared to~\cite{prat_5}.
A configuration option, which we refer to as~LG, 
is used to decide whether the LSTM is stacked on top of the
GRU ($\mbox{LG} = \mbox{false}$ in this case) or GRU is stacked on 
top of the LSTM ($\mbox{LG} = \mbox{true}$). 
For example, when LG is ``true,'' opcode inputs are first passed 
to LSTM layers, with the output of the LSTM (i.e. the hidden cells) becoming
input to the GRU layers. The output of the GRU is then
passed to fully connected layers that are used to classify the input data. 
We have applied a dropout layer with~0.3 probability for models with more 
than one layer.

\begin{figure}[!htb]
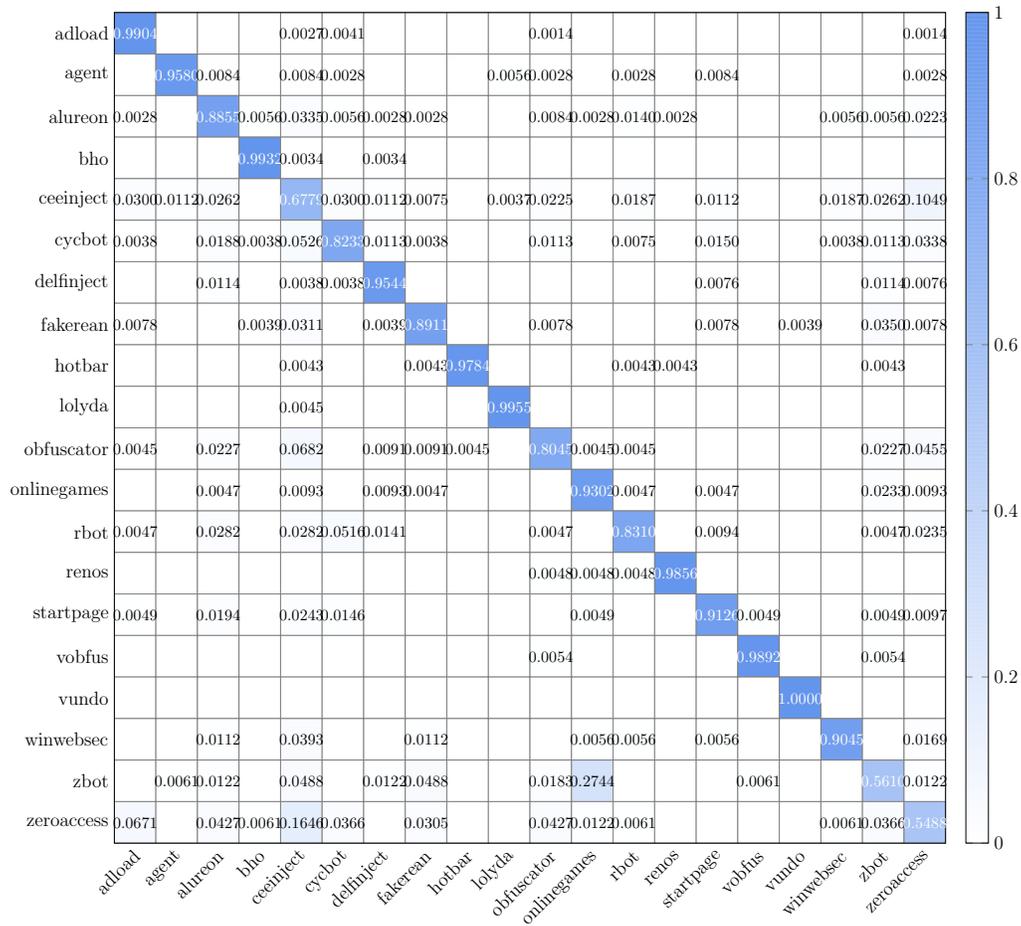

	\centering
	\input figures/conf_StackedMalware_Model1.tex
	\caption{Confusion matrix for stacked LSTM-GRU experiment}
	\label{fig:conf_conf_StackedMalware_Model1}
\end{figure}

The best overall accuracy we obtain for our stacked 
LSTM-GRU experiments is~0.8990; the confusion
matrix for this case is given in Figure~\ref{fig:conf_conf_StackedMalware_Model1}.
This is somewhat disappointing,
as it is in between the
results obtained for our LSTM and GRU models.

\subsection{Transfer Learning}

Finally, we have considered two popular image-based transfer learning
models, namely RestNet152 and VGG-19. These are
models that have been pre-trained on extremely large image datasets,
and we simply retrain the last few layers for the malware dataset
under consideration in this paper, while the earlier layers are frozen during training.
The parameters used in these experiments are 
summarized in Table~\ref{tab:classRN}.

\begin{table}[!htb]
\caption{Transfer learning model parameters}\label{tab:classRN}
\centering
{\small
	\begin{tabular}{c|c|c|c|c}\midrule\midrule
		\multirow{2}{*}{\textbf{Classifier}} 
			& \multirow{2}{*}{\textbf{Hyperparameter}} 
				& \multirow{2}{*}{\textbf{Tested values}} 
				& \multicolumn{2}{c}{\textbf{Accuracy}} \\
			& & & \textbf{Train} & \textbf{Test} \\  
				\midrule
		\multirow{4}{*}{ResNet152} 
			& \texttt{image\un dim} & $256$ 
				& \multirow{4}{*} {$0.9811$} & \multirow{4}{*} {$0.9150$} \\ 
			& \texttt{learning\un rate} & $[0.001,\textbf{0.0001}]$ & \\
			& \texttt{batch\un size} & $256$ & \\
			& \texttt{epochs} & $20$ & \\
				\midrule
		\multirow{4}{*}{VGG-19} 
			& \texttt{image\un dim} & $256$ 
				& \multirow{4}{*} {$0.9690$} & \multirow{4}{*} {$0.9216$} \\ 
			& \texttt{learning\un rate} & $[0.001,\textbf{0.0001}]$ & \\
			& \texttt{batch\un size} & $256$ & \\
			& \texttt{epochs} & $20$ & \\
				\midrule\midrule
	\end{tabular}
}
\end{table}

For ResNet152, the model parameters for layer four were unfrozen for training. 
We also added two more layers of fully connected neurons 
for training. Resnet152 is pre-trained based on~1000 classes and 
hence its last fully connected layer has output dimensions of~1000. 
We reduce this output dimension to~500 via another fully connected layer, 
and an additional fully connected layer further reduces the output dimension 
to~20, which is the number of classes in our dataset. 

For VGG-19, we froze all layers except~34, 35, and~36. 
As with ResNet152, we added two more layers of fully connected 
neurons to reduce the output dimension from~1000 to~20.

For all of our transfer learning experiments, we use a batch size 
of~256 and trained each model for~20 epochs with 
learning rates of~0.001 and~0.0001.
Both ResNet152 and VGG-19 expect image dimensions of~$224\times 224$
and hence we resize our~$256\times 256$ images 
to~$224\times 224$.

The performance of these transfer learning models was the best
of our deep learning experiments, with ResNet152 achieving
an overall accuracy of~0.9150 and VGG-19 doing slightly better
at~0.9216. The confusion matrix for VGG-19 is given in Figure~\ref{fig:conf_vgg19};
we omit the confusion matrix for ResNet152 since it is similar, but
marginally worse. As compared to the other image-based 
deep learning models we have considered,
we see marked improvement in the classification accuracy of the most
challenging families, such as Obfuscator.

\subsection{Discussion}

\begin{figure}[!htb]
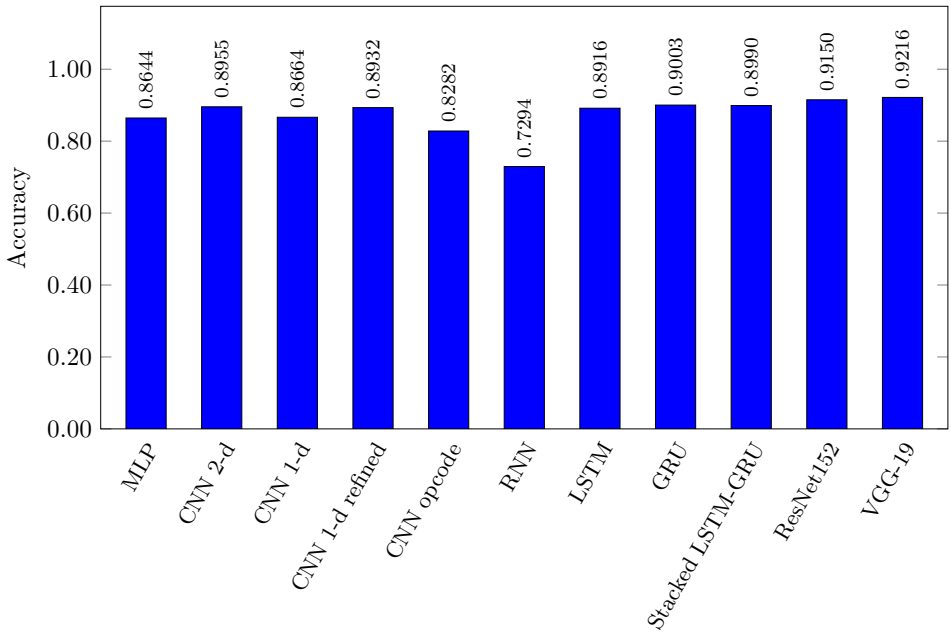

	\centering
	\input figures/bar.tex
	\caption{Comparison of results}\label{fig:bar}
\end{figure}

The results of the malware classification experiments 
discussed in this section are summarized in Figure~\ref{fig:bar}.
We see that among the deep learning techniques, the image-based
pre-trained models, namely, ResNet152 and VGG-19, perform 
best, with VGG-19 classifying more than~92\%\ of the samples
correctly. The best of our other (i.e., not pre-trained) image-based 
models achieved slightly less than~90\%\ accuracy.

Although the opcode-based results performed relatively poorly overall,
it is interesting to note that they were able to classify some families
with higher accuracy than any of the image-based models.
This suggests that a model that combines both
image features and opcode features might be more effective
than either approach individually.

%We also note that the classic machine learning techniques can 
%produce higher accuracies. However, this is misleading, as all
%of these techniques suffer from obvious signs of overfitting.

\section{Conclusions and Future Work}\label{sect:con}

Malware classification is a fundamental and challenging problem in 
information security. Previous work has indicated that treating malware
executables as images and applying image-based techniques 
can yield strong classification results. 

In this paper, we provided results from a vast number of learning
experiments, comparing deep learning techniques using
image-based features to some cases involving opcode features.
For our
deep learning techniques, we focused on multilayer perceptrons (MLP), 
convolutional neural networks (CNN), and recurrent neural networks (RNN),
including long short-term memory (LSTM) and gated recurrent units (GRU).
We also experimented with the image-based transfer learning 
techniques ResNet152 and VGG-19.
Among these
techniques, the image-based transfer learning models performed
the best, with the best classification accuracy exceeding~92\%.

For future work, additional transfer learning experiments would
be worthwhile, as there are many more parameters that could be tested.
Larger and more diverse datasets could be considered.
In addition, it would be interesting to consider both image-based
and opcode features as part of a combined classification technique.
As noted above, the opcode-based
techniques perform worse overall, but they do provide better results
for some families that are particularly challenging to distinguish
based only on image features.

\bibliographystyle{plain}

\bibliography{other.bib,references.bib,Stamp-Mark.bib}

\clearpage

\appendix

\section*{Appendix: Confusion Matrices}

\begin{figure}[!htb]
	\centering
	\input figures/conf_CNNMalware_Model1.tex
	\caption{Confusion matrix for CNN 2-d experiment}
	\label{fig:conf_CNN_small}
\end{figure}

\begin{figure}[!htb]
	\centering
	\input figures/conf_CNNMalware_Model4.tex
	\caption{Confusion matrix for CNN 1-d experiment}
	\label{fig:conf_CNN_1d}
\end{figure}

\begin{figure}[!htb]
	\centering
	\input figures/conf_tl_vgg19.tex
	\caption{Confusion matrix for VGG-19 experiment}
	\label{fig:conf_vgg19}
\end{figure}

%%%%% Where are vanilla RNN, LSTM, and ResNet152 confusion matrices ?????

\end{document}

%% file: preamble.tex
\usepackage{amsmath, amsfonts, amssymb, amsxtra, amsopn}
\usepackage{pgfplots}
\pgfplotsset{compat=1.13}
\usepackage{pgfplotstable}
\usepackage{graphicx}
\usepackage{multirow}
\usepackage{booktabs}
\usepackage{listings}
\usepackage{cmap}
\usepackage{colortbl}
\usepackage{adjustbox}
\usepackage{epsfig}

\usepackage[tableposition=top]{caption}

\PassOptionsToPackage{hyphens}{url}

\usepackage{hyperref}
\hypersetup{colorlinks=true,linkcolor=black,citecolor=black,urlcolor=blue,filecolor=black}
\hypersetup{pdfpagemode=UseNone,pdfstartview=}

\usepackage{enumitem}
\setlist[itemize]{noitemsep, topsep=0pt}

\advance\oddsidemargin by -0.2in
\advance\textwidth by 0.4in

\advance\topmargin by -0.25in
\advance\textheight by 0.5in

\long\def\symbolfootnotetext[#1]#2{\begingroup%
\def\thefootnote{\fnsymbol{footnote}}\footnotetext[#1]{#2}\endgroup}

\def\zz{\phantom{0}}

\def\un{\underline{\phantom{M}}}

%% file: figures/bar.tex
\begin{tikzpicture}[scale=0.95, every node/.style={scale=1.0}]
\begin{axis}[%bar shift=0pt,
        width  = 0.85*\textwidth,
        height = 7.5cm,
        ymin=0.0,ymax=1.175,
        ytick={0.0,0.2,0.4,0.6,0.8,1.0},
        major x tick style = transparent,
        ybar=5*\pgflinewidth,
        bar width=16.0pt,
%        ymajorgrids = true,
        ylabel = {Accuracy},
        symbolic x coords={MLP, CNN 2-d, CNN 1-d, CNN 1-d refined, CNN opcode, RNN, LSTM, GRU, Stacked LSTM-GRU, ResNet152, VGG-19},
	y tick label style={
    		/pgf/number format/.cd,
   		fixed,
   		fixed zerofill,
    		precision=2},
%	yticklabel pos=right,
        xtick = data,
        x tick label style={
        		rotate=60,
		font=\small,
		anchor=north east,
		inner sep=0mm},
%		font=\small},
%        scaled y ticks = false,
	%%%%% numbers on bars and rotated
        nodes near coords,
        every node near coord/.append style={rotate=90, 
        								   anchor=west,
								   font=\footnotesize,
								   /pgf/number format/.cd,
								   fixed,
								   fixed zerofill,
								   precision=4},
        %%%%%
%        enlarge x limits=0.03,
        enlarge x limits=0.06,
        legend cell align=left,
        legend style={
%                at={(1,1.05)},
%                anchor=south east,
%	        nodes={rotate=90},%%%%% rotate text in legend
%                at={(0.125,0)},
%                at={(0.125,0)},
%                at={(0.8775,0)},
                at={(0.89,0.02)},
                anchor=south,
                column sep=1ex
        },
]
\addplot [fill=blue,opacity=1.00]
coordinates {
(MLP, 0.8644)
(CNN 2-d, 0.8955)
(CNN 1-d, 0.8664)
(CNN 1-d refined, 0.8932)
(CNN opcode, 0.8282)
(RNN, 0.7294)
(LSTM, 0.8916)
(GRU, 0.9003)
(Stacked LSTM-GRU, 0.8990)
(ResNet152, 0.9150)
(VGG-19, 0.9216)
};
\end{axis}
\end{tikzpicture}

%% file: figures/conf_ANNMalware_Model2.tex
%\begin{tikzpicture}[scale=0.60,every node/.style={scale=0.9}]
\begin{tikzpicture}[scale=0.6]
    \begin{axis}[%colorbar/width=2.5mm,
        width=20cm,
        height=20cm,
%        colormap={blackwhite}{gray(0cm)=(1); gray(1cm)=(0.5)},
%   colormap={bluewhite}{color=(white) color=(blue)},
%   colormap={bluewhite}{color=(white) rgb255=(0,191,255)},
    colormap={bluewhite}{color=(white) rgb255=(100,149,237)},
        xticklabels={adload,agent,alureon,bho,ceeinject,cycbot,delfinject,fakerean,hotbar,lolyda,obfuscator,onlinegames,rbot,renos,startpage,vobfus,vundo,winwebsec,zbot,zeroaccess},
        xtick={0,...,19},
        xtick style={draw=none},
    xticklabel style={anchor=east,rotate=45,yshift=-5pt},
        yticklabels={adload,agent,alureon,bho,ceeinject,cycbot,delfinject,fakerean,hotbar,lolyda,obfuscator,onlinegames,rbot,renos,startpage,vobfus,vundo,winwebsec,zbot,zeroaccess},
        ytick={0,...,19},
        ytick style={draw=none},
        enlargelimits=false,
        xticklabel style={font=\large},
        yticklabel style={font=\large},
        colorbar,
        colorbar style={
%           width=0.05*\pgfkeysvalueof{/pgfplots/parent axis width},%%% added this
%           height=0.5*\pgfkeysvalueof{/pgfplots/parent axis height},
%       plot graphics/node/.style={scale=1.33,anchor=south west,inner sep=0pt,}, %%% scale colorbar fill %%%
            ytick={0.0,0.2,0.4,0.6,0.8,1.0},
            yticklabels={0.0,0.2,0.4,0.6,0.8,1.0},
            yticklabel={\pgfmathprintnumber\tick},
            yticklabel style={font=\large,
                    /pgf/number format/fixed,
            /pgf/number format/precision=1}
        },
%        point meta min=0,
%        point meta max=100,
        point meta min=0.0,
        point meta max=1.0,
        nodes near coords={\pgfmathprintnumber\pgfplotspointmeta},
        % ---------------------------------------------------------------------
        % show `nodes near coords' but adapt the style so that values
        % above a threshold get another style
        % (adapted from <http://tex.stackexchange.com/a/141006/95441>)
        % #1: the THRESHOLD after which we switch to a special display.
        nodes near coords black white/.style={
            % define the style of the nodes with "small" values
            small value/.style={
                yshift=-7pt,
%                text=white,
                text=black,
                /pgf/number format/fixed,
                /pgf/number format/precision=4,
                /pgf/number format/zerofill
            },
            % define the style of the nodes with "large" values
            large value/.style={
                yshift=-7pt,
%                text=black,
                text=white,
                /pgf/number format/fixed,
                /pgf/number format/precision=4,
                /pgf/number format/zerofill
            },
            every node near coord/.style={
                check for zero/.code={
                    \pgfmathfloatifflags{\pgfplotspointmeta}{0}{
                        % If meta=0, make the node a coordinate
                        % (which doesn't have text)
                        \pgfkeys{/tikz/coordinate}
                    }{
                        \begingroup
                        % this group is merely to switch to FPU locally.
                        % Might be unnecessary, but who knows.
                        \pgfkeys{/pgf/fpu}
                        \pgfmathparse{\pgfplotspointmeta<#1}
                        \global\let\result=\pgfmathresult
                        \endgroup
                        %
                        % simplifies debugging:
                        %\show\result
                        %
                        \pgfmathfloatcreate{1}{1.0}{0}
                        \let\ONE=\pgfmathresult
                        \ifx\result\ONE
                            % AH: our condition 'y < #1' is met.
                            \pgfkeysalso{/pgfplots/small value}
                        \else
                            % ok, proceed as usual.
                            \pgfkeysalso{/pgfplots/large value}
                        \fi
                    }
                },
                check for zero,
            },
        },
        % asign a value to the new style which is the threshold at which
        % the two style `small value' or `large value' are used
%        nodes near coords black white=50,
        nodes near coords black white=0.5,
        % -----------------------------------------------------------------
    ]
        \addplot[
            matrix plot,
            mesh/cols=20,
            point meta=explicit,draw=gray
        ] table [meta=C] {
            x y C
0 0 0.9821
1 0 0.0
2 0 0.0
3 0 0.0
4 0 0.0
5 0 0.0
6 0 0.0045
7 0 0.0
8 0 0.0
9 0 0.0
10 0 0.009
11 0 0.0
12 0 0.0045
13 0 0.0
14 0 0.0
15 0 0.0
16 0 0.0
17 0 0.0
18 0 0.0
19 0 0.0
0 1 0.0056
1 1 0.7778
2 1 0.0111
3 1 0.0
4 1 0.0056
5 1 0.0056
6 1 0.0056
7 1 0.0056
8 1 0.0
9 1 0.0
10 1 0.05
11 1 0.0889
12 1 0.0222
13 1 0.0056
14 1 0.0056
15 1 0.0
16 1 0.0056
17 1 0.0
18 1 0.0056
19 1 0.0
0 2 0.0039
1 2 0.0117
2 2 0.8164
3 2 0.0039
4 2 0.0039
5 2 0.0039
6 2 0.0
7 2 0.0039
8 2 0.0039
9 2 0.0039
10 2 0.0312
11 2 0.0234
12 2 0.0078
13 2 0.0117
14 2 0.0
15 2 0.0195
16 2 0.0391
17 2 0.0039
18 2 0.0039
19 2 0.0039
0 3 0.0041
1 3 0.0041
2 3 0.0041
3 3 0.9506
4 3 0.0
5 3 0.0
6 3 0.0
7 3 0.0
8 3 0.0041
9 3 0.0
10 3 0.0041
11 3 0.0
12 3 0.0041
13 3 0.0041
14 3 0.0082
15 3 0.0
16 3 0.0082
17 3 0.0
18 3 0.0041
19 3 0.0
0 4 0.0
1 4 0.0222
2 4 0.0056
3 4 0.0111
4 4 0.85
5 4 0.0
6 4 0.0111
7 4 0.0
8 4 0.0
9 4 0.0
10 4 0.0556
11 4 0.0111
12 4 0.0056
13 4 0.0
14 4 0.0
15 4 0.0
16 4 0.0222
17 4 0.0056
18 4 0.0
19 4 0.0
0 5 0.0
1 5 0.0
2 5 0.0049
3 5 0.0
4 5 0.0
5 5 0.9557
6 5 0.0099
7 5 0.0
8 5 0.0
9 5 0.0
10 5 0.0148
11 5 0.0049
12 5 0.0049
13 5 0.0
14 5 0.0
15 5 0.0
16 5 0.0049
17 5 0.0
18 5 0.0
19 5 0.0
0 6 0.0042
1 6 0.0792
2 6 0.0083
3 6 0.0042
4 6 0.025
5 6 0.0083
6 6 0.5625
7 6 0.0208
8 6 0.0125
9 6 0.0
10 6 0.1167
11 6 0.0083
12 6 0.075
13 6 0.0125
14 6 0.0042
15 6 0.0
16 6 0.0375
17 6 0.0042
18 6 0.0042
19 6 0.0125
0 7 0.0
1 7 0.01
2 7 0.0
3 7 0.0
4 7 0.01
5 7 0.0
6 7 0.025
7 7 0.865
8 7 0.0
9 7 0.0
10 7 0.035
11 7 0.0
12 7 0.025
13 7 0.005
14 7 0.005
15 7 0.0
16 7 0.0
17 7 0.005
18 7 0.005
19 7 0.01
0 8 0.0
1 8 0.0
2 8 0.0
3 8 0.0035
4 8 0.0
5 8 0.0
6 8 0.0
7 8 0.0
8 8 0.986
9 8 0.0
10 8 0.0035
11 8 0.0
12 8 0.007
13 8 0.0
14 8 0.0
15 8 0.0
16 8 0.0
17 8 0.0
18 8 0.0
19 8 0.0
0 9 0.0
1 9 0.0
2 9 0.0
3 9 0.0
4 9 0.0
5 9 0.0
6 9 0.0
7 9 0.0
8 9 0.0
9 9 1.0
10 9 0.0
11 9 0.0
12 9 0.0
13 9 0.0
14 9 0.0
15 9 0.0
16 9 0.0
17 9 0.0
18 9 0.0
19 9 0.0
0 10 0.0
1 10 0.0526
2 10 0.0526
3 10 0.0105
4 10 0.014
5 10 0.014
6 10 0.0421
7 10 0.0491
8 10 0.0035
9 10 0.0035
10 10 0.5474
11 10 0.014
12 10 0.0982
13 10 0.0105
14 10 0.0175
15 10 0.0
16 10 0.0526
17 10 0.0035
18 10 0.0105
19 10 0.0035
0 11 0.0
1 11 0.0276
2 11 0.0
3 11 0.0079
4 11 0.0039
5 11 0.0039
6 11 0.0039
7 11 0.0079
8 11 0.0
9 11 0.0079
10 11 0.0236
11 11 0.8898
12 11 0.0079
13 11 0.0
14 11 0.0
15 11 0.0
16 11 0.0079
17 11 0.0039
18 11 0.0039
19 11 0.0
0 12 0.0
1 12 0.0248
2 12 0.005
3 12 0.0
4 12 0.005
5 12 0.0099
6 12 0.0545
7 12 0.005
8 12 0.005
9 12 0.0
10 12 0.0792
11 12 0.0
12 12 0.7574
13 12 0.005
14 12 0.0
15 12 0.005
16 12 0.0248
17 12 0.0
18 12 0.005
19 12 0.0149
0 13 0.0
1 13 0.0038
2 13 0.0038
3 13 0.0
4 13 0.0038
5 13 0.0
6 13 0.0075
7 13 0.0075
8 13 0.0
9 13 0.0
10 13 0.0263
11 13 0.0038
12 13 0.0338
13 13 0.8722
14 13 0.0113
15 13 0.0
16 13 0.0113
17 13 0.0038
18 13 0.0113
19 13 0.0
0 14 0.0
1 14 0.0961
2 14 0.0044
3 14 0.0044
4 14 0.0044
5 14 0.0
6 14 0.0175
7 14 0.0087
8 14 0.0
9 14 0.0
10 14 0.0131
11 14 0.0044
12 14 0.0131
13 14 0.0044
14 14 0.8297
15 14 0.0
16 14 0.0
17 14 0.0
18 14 0.0
19 14 0.0
0 15 0.0
1 15 0.0226
2 15 0.0113
3 15 0.0
4 15 0.0
5 15 0.0
6 15 0.0056
7 15 0.0
8 15 0.0
9 15 0.0
10 15 0.0226
11 15 0.0
12 15 0.0113
13 15 0.0
14 15 0.0
15 15 0.9209
16 15 0.0056
17 15 0.0
18 15 0.0
19 15 0.0
0 16 0.0027
1 16 0.0164
2 16 0.0219
3 16 0.0027
4 16 0.0055
5 16 0.0
6 16 0.0137
7 16 0.0027
8 16 0.0
9 16 0.0
10 16 0.0301
11 16 0.0027
12 16 0.0274
13 16 0.0055
14 16 0.0055
15 16 0.0
16 16 0.8521
17 16 0.0027
18 16 0.0082
19 16 0.0
0 17 0.0
1 17 0.0
2 17 0.0
3 17 0.0
4 17 0.0
5 17 0.0
6 17 0.0
7 17 0.0
8 17 0.0
9 17 0.0
10 17 0.0
11 17 0.0
12 17 0.0
13 17 0.0
14 17 0.0
15 17 0.0
16 17 0.0
17 17 1.0
18 17 0.0
19 17 0.0
0 18 0.0
1 18 0.0
2 18 0.0028
3 18 0.0028
4 18 0.0
5 18 0.0
6 18 0.0028
7 18 0.0028
8 18 0.0
9 18 0.0
10 18 0.0085
11 18 0.0
12 18 0.0
13 18 0.0028
14 18 0.0
15 18 0.0
16 18 0.0057
17 18 0.0
18 18 0.9688
19 18 0.0028
0 19 0.0
1 19 0.0
2 19 0.0
3 19 0.0
4 19 0.0
5 19 0.0
6 19 0.0042
7 19 0.0042
8 19 0.0
9 19 0.0
10 19 0.0
11 19 0.0
12 19 0.0
13 19 0.0042
14 19 0.0
15 19 0.0
16 19 0.0
17 19 0.0
18 19 0.0
19 19 0.9875
        };
    \end{axis}
\end{tikzpicture}
%
%\caption{I'm confused~5!}\label{tab:CM5}
%\end{figure*}

%% file: figures/conf_CNNMalware_Model1.tex
%\begin{tikzpicture}[scale=0.60,every node/.style={scale=0.9}]
\begin{tikzpicture}[scale=0.6]
    \begin{axis}[%colorbar/width=2.5mm,
        width=20cm,
        height=20cm,
%        colormap={blackwhite}{gray(0cm)=(1); gray(1cm)=(0.5)},
%   colormap={bluewhite}{color=(white) color=(blue)},
%   colormap={bluewhite}{color=(white) rgb255=(0,191,255)},
    colormap={bluewhite}{color=(white) rgb255=(100,149,237)},
        xticklabels={adload,agent,alureon,bho,ceeinject,cycbot,delfinject,fakerean,hotbar,lolyda,obfuscator,onlinegames,rbot,renos,startpage,vobfus,vundo,winwebsec,zbot,zeroaccess},
        xtick={0,...,19},
        xtick style={draw=none},
    xticklabel style={anchor=east,rotate=45,yshift=-5pt},
        yticklabels={adload,agent,alureon,bho,ceeinject,cycbot,delfinject,fakerean,hotbar,lolyda,obfuscator,onlinegames,rbot,renos,startpage,vobfus,vundo,winwebsec,zbot,zeroaccess},
        ytick={0,...,19},
        ytick style={draw=none},
        enlargelimits=false,
        xticklabel style={font=\large},
        yticklabel style={font=\large},
        colorbar,
        colorbar style={
%           width=0.05*\pgfkeysvalueof{/pgfplots/parent axis width},%%% added this
%           height=0.5*\pgfkeysvalueof{/pgfplots/parent axis height},
%       plot graphics/node/.style={scale=1.33,anchor=south west,inner sep=0pt,}, %%% scale colorbar fill %%%
            ytick={0.0,0.2,0.4,0.6,0.8,1.0},
            yticklabels={0.0,0.2,0.4,0.6,0.8,1.0},
            yticklabel={\pgfmathprintnumber\tick},
            yticklabel style={font=\large,
                    /pgf/number format/fixed,
            /pgf/number format/precision=1}
        },
%        point meta min=0,
%        point meta max=100,
        point meta min=0.0,
        point meta max=1.0,
        nodes near coords={\pgfmathprintnumber\pgfplotspointmeta},
        % ---------------------------------------------------------------------
        % show `nodes near coords' but adapt the style so that values
        % above a threshold get another style
        % (adapted from <http://tex.stackexchange.com/a/141006/95441>)
        % #1: the THRESHOLD after which we switch to a special display.
        nodes near coords black white/.style={
            % define the style of the nodes with "small" values
            small value/.style={
                yshift=-7pt,
%                text=white,
                text=black,
                /pgf/number format/fixed,
                /pgf/number format/precision=4,
                /pgf/number format/zerofill
            },
            % define the style of the nodes with "large" values
            large value/.style={
                yshift=-7pt,
%                text=black,
                text=white,
                /pgf/number format/fixed,
                /pgf/number format/precision=4,
                /pgf/number format/zerofill
            },
            every node near coord/.style={
                check for zero/.code={
                    \pgfmathfloatifflags{\pgfplotspointmeta}{0}{
                        % If meta=0, make the node a coordinate
                        % (which doesn't have text)
                        \pgfkeys{/tikz/coordinate}
                    }{
                        \begingroup
                        % this group is merely to switch to FPU locally.
                        % Might be unnecessary, but who knows.
                        \pgfkeys{/pgf/fpu}
                        \pgfmathparse{\pgfplotspointmeta<#1}
                        \global\let\result=\pgfmathresult
                        \endgroup
                        %
                        % simplifies debugging:
                        %\show\result
                        %
                        \pgfmathfloatcreate{1}{1.0}{0}
                        \let\ONE=\pgfmathresult
                        \ifx\result\ONE
                            % AH: our condition 'y < #1' is met.
                            \pgfkeysalso{/pgfplots/small value}
                        \else
                            % ok, proceed as usual.
                            \pgfkeysalso{/pgfplots/large value}
                        \fi
                    }
                },
                check for zero,
            },
        },
        % asign a value to the new style which is the threshold at which
        % the two style `small value' or `large value' are used
%        nodes near coords black white=50,
        nodes near coords black white=0.5,
        % -----------------------------------------------------------------
    ]
        \addplot[
            matrix plot,
            mesh/cols=20,
            point meta=explicit,draw=gray
        ] table [meta=C] {
            x y C
			0 0 0.9674
			1 0 0.0047
			2 0 0.0
			3 0 0.0047
			4 0 0.0
			5 0 0.0
			6 0 0.0047
			7 0 0.0
			8 0 0.0
			9 0 0.0
			10 0 0.0186
			11 0 0.0
			12 0 0.0
			13 0 0.0
			14 0 0.0
			15 0 0.0
			16 0 0.0
			17 0 0.0
			18 0 0.0
			19 0 0.0
			0 1 0.0
			1 1 0.7765
			2 1 0.0056
			3 1 0.0
			4 1 0.0
			5 1 0.0112
			6 1 0.0056
			7 1 0.0
			8 1 0.0
			9 1 0.0
			10 1 0.0615
			11 1 0.067
			12 1 0.0112
			13 1 0.0168
			14 1 0.0056
			15 1 0.0056
			16 1 0.0335
			17 1 0.0
			18 1 0.0
			19 1 0.0
			0 2 0.0
			1 2 0.0037
			2 2 0.8764
			3 2 0.0
			4 2 0.0
			5 2 0.0
			6 2 0.0
			7 2 0.0
			8 2 0.0
			9 2 0.0
			10 2 0.0375
			11 2 0.0075
			12 2 0.015
			13 2 0.0037
			14 2 0.0
			15 2 0.0037
			16 2 0.0524
			17 2 0.0
			18 2 0.0
			19 2 0.0
			0 3 0.0045
			1 3 0.0
			2 3 0.0
			3 3 0.9865
			4 3 0.0
			5 3 0.0
			6 3 0.0
			7 3 0.0
			8 3 0.0
			9 3 0.0
			10 3 0.0045
			11 3 0.0
			12 3 0.0
			13 3 0.0
			14 3 0.0
			15 3 0.0
			16 3 0.0045
			17 3 0.0
			18 3 0.0
			19 3 0.0
			0 4 0.0
			1 4 0.0057
			2 4 0.0229
			3 4 0.0
			4 4 0.8286
			5 4 0.0114
			6 4 0.0
			7 4 0.0
			8 4 0.0
			9 4 0.0
			10 4 0.0571
			11 4 0.0
			12 4 0.0171
			13 4 0.0057
			14 4 0.0057
			15 4 0.0
			16 4 0.04
			17 4 0.0
			18 4 0.0
			19 4 0.0057
			0 5 0.0
			1 5 0.0
			2 5 0.0097
			3 5 0.0
			4 5 0.0
			5 5 0.9565
			6 5 0.0
			7 5 0.0
			8 5 0.0
			9 5 0.0
			10 5 0.0048
			11 5 0.0
			12 5 0.0145
			13 5 0.0048
			14 5 0.0
			15 5 0.0
			16 5 0.0097
			17 5 0.0
			18 5 0.0
			19 5 0.0
			0 6 0.0
			1 6 0.0205
			2 6 0.0205
			3 6 0.0
			4 6 0.0
			5 6 0.0041
			6 6 0.7131
			7 6 0.0041
			8 6 0.0
			9 6 0.0
			10 6 0.1107
			11 6 0.0041
			12 6 0.0205
			13 6 0.0369
			14 6 0.0328
			15 6 0.0
			16 6 0.0287
			17 6 0.0
			18 6 0.0041
			19 6 0.0
			0 7 0.0
			1 7 0.0
			2 7 0.037
			3 7 0.0
			4 7 0.0
			5 7 0.0139
			6 7 0.0093
			7 7 0.8333
			8 7 0.0
			9 7 0.0
			10 7 0.0417
			11 7 0.0
			12 7 0.037
			13 7 0.0
			14 7 0.0
			15 7 0.0
			16 7 0.0278
			17 7 0.0
			18 7 0.0
			19 7 0.0
			0 8 0.0
			1 8 0.0
			2 8 0.0066
			3 8 0.0
			4 8 0.0
			5 8 0.0
			6 8 0.0
			7 8 0.0
			8 8 0.9801
			9 8 0.0
			10 8 0.0033
			11 8 0.0
			12 8 0.0033
			13 8 0.0
			14 8 0.0033
			15 8 0.0
			16 8 0.0033
			17 8 0.0
			18 8 0.0
			19 8 0.0
			0 9 0.0
			1 9 0.0
			2 9 0.0051
			3 9 0.0
			4 9 0.0
			5 9 0.0
			6 9 0.0051
			7 9 0.0
			8 9 0.0
			9 9 0.9847
			10 9 0.0
			11 9 0.0
			12 9 0.0
			13 9 0.0
			14 9 0.0
			15 9 0.0
			16 9 0.0051
			17 9 0.0
			18 9 0.0
			19 9 0.0
			0 10 0.0
			1 10 0.0251
			2 10 0.0681
			3 10 0.0
			4 10 0.0108
			5 10 0.0072
			6 10 0.0394
			7 10 0.0251
			8 10 0.0
			9 10 0.0
			10 10 0.6201
			11 10 0.0179
			12 10 0.0538
			13 10 0.0143
			14 10 0.0072
			15 10 0.0036
			16 10 0.1004
			17 10 0.0
			18 10 0.0072
			19 10 0.0
			0 11 0.0037
			1 11 0.0149
			2 11 0.0149
			3 11 0.0
			4 11 0.0
			5 11 0.0037
			6 11 0.0149
			7 11 0.0
			8 11 0.0
			9 11 0.0037
			10 11 0.0112
			11 11 0.881
			12 11 0.0074
			13 11 0.0074
			14 11 0.0
			15 11 0.0037
			16 11 0.0335
			17 11 0.0
			18 11 0.0
			19 11 0.0
			0 12 0.0
			1 12 0.015
			2 12 0.01
			3 12 0.005
			4 12 0.0
			5 12 0.015
			6 12 0.015
			7 12 0.005
			8 12 0.0
			9 12 0.0
			10 12 0.095
			11 12 0.005
			12 12 0.76
			13 12 0.01
			14 12 0.0
			15 12 0.01
			16 12 0.05
			17 12 0.0
			18 12 0.005
			19 12 0.0
			0 13 0.0
			1 13 0.0
			2 13 0.0177
			3 13 0.0
			4 13 0.0
			5 13 0.0044
			6 13 0.0133
			7 13 0.0
			8 13 0.0
			9 13 0.0
			10 13 0.0265
			11 13 0.0
			12 13 0.0044
			13 13 0.885
			14 13 0.0
			15 13 0.0
			16 13 0.0487
			17 13 0.0
			18 13 0.0
			19 13 0.0
			0 14 0.0
			1 14 0.055
			2 14 0.005
			3 14 0.0
			4 14 0.0
			5 14 0.01
			6 14 0.015
			7 14 0.0
			8 14 0.0
			9 14 0.0
			10 14 0.045
			11 14 0.0
			12 14 0.015
			13 14 0.01
			14 14 0.835
			15 14 0.01
			16 14 0.0
			17 14 0.0
			18 14 0.0
			19 14 0.0
			0 15 0.0
			1 15 0.0056
			2 15 0.0056
			3 15 0.0
			4 15 0.0
			5 15 0.0056
			6 15 0.0056
			7 15 0.0
			8 15 0.0
			9 15 0.0
			10 15 0.0056
			11 15 0.0
			12 15 0.0056
			13 15 0.0
			14 15 0.0
			15 15 0.9435
			16 15 0.0226
			17 15 0.0
			18 15 0.0
			19 15 0.0
			0 16 0.003
			1 16 0.0061
			2 16 0.0061
			3 16 0.0
			4 16 0.0
			5 16 0.003
			6 16 0.0
			7 16 0.0
			8 16 0.0
			9 16 0.0
			10 16 0.0274
			11 16 0.0091
			12 16 0.0274
			13 16 0.0122
			14 16 0.0
			15 16 0.0091
			16 16 0.8936
			17 16 0.0
			18 16 0.003
			19 16 0.0
			0 17 0.0
			1 17 0.0
			2 17 0.0
			3 17 0.0
			4 17 0.0
			5 17 0.0
			6 17 0.0
			7 17 0.0026
			8 17 0.0
			9 17 0.0
			10 17 0.0026
			11 17 0.0
			12 17 0.0
			13 17 0.0
			14 17 0.0
			15 17 0.0
			16 17 0.0013
			17 17 0.9936
			18 17 0.0
			19 17 0.0
			0 18 0.0
			1 18 0.0
			2 18 0.0082
			3 18 0.0
			4 18 0.0
			5 18 0.0
			6 18 0.0
			7 18 0.0
			8 18 0.0
			9 18 0.0
			10 18 0.0137
			11 18 0.0027
			12 18 0.0027
			13 18 0.0
			14 18 0.0
			15 18 0.0
			16 18 0.0164
			17 18 0.0
			18 18 0.9534
			19 18 0.0027
			0 19 0.0
			1 19 0.0
			2 19 0.0043
			3 19 0.0
			4 19 0.0
			5 19 0.0043
			6 19 0.0
			7 19 0.0
			8 19 0.0
			9 19 0.0
			10 19 0.0
			11 19 0.0043
			12 19 0.0
			13 19 0.0
			14 19 0.0
			15 19 0.0
			16 19 0.0
			17 19 0.0
			18 19 0.0
			19 19 0.9872

       };
    \end{axis}
\end{tikzpicture}
%
%\caption{I'm confused~5!}\label{tab:CM5}
%\end{figure*}

%% file: figures/conf_CNNMalware_Model4.tex
%\begin{tikzpicture}[scale=0.60,every node/.style={scale=0.9}]
\begin{tikzpicture}[scale=0.6]
    \begin{axis}[%colorbar/width=2.5mm,
        width=20cm,
        height=20cm,
%        colormap={blackwhite}{gray(0cm)=(1); gray(1cm)=(0.5)},
%   colormap={bluewhite}{color=(white) color=(blue)},
%   colormap={bluewhite}{color=(white) rgb255=(0,191,255)},
    colormap={bluewhite}{color=(white) rgb255=(100,149,237)},
        xticklabels={adload,agent,alureon,bho,ceeinject,cycbot,delfinject,fakerean,hotbar,lolyda,obfuscator,onlinegames,rbot,renos,startpage,vobfus,vundo,winwebsec,zbot,zeroaccess},
        xtick={0,...,19},
        xtick style={draw=none},
    xticklabel style={anchor=east,rotate=45,yshift=-5pt},
        yticklabels={adload,agent,alureon,bho,ceeinject,cycbot,delfinject,fakerean,hotbar,lolyda,obfuscator,onlinegames,rbot,renos,startpage,vobfus,vundo,winwebsec,zbot,zeroaccess},
        ytick={0,...,19},
        ytick style={draw=none},
        enlargelimits=false,
        xticklabel style={font=\large},
        yticklabel style={font=\large},
        colorbar,
        colorbar style={
%           width=0.05*\pgfkeysvalueof{/pgfplots/parent axis width},%%% added this
%           height=0.5*\pgfkeysvalueof{/pgfplots/parent axis height},
%       plot graphics/node/.style={scale=1.33,anchor=south west,inner sep=0pt,}, %%% scale colorbar fill %%%
            ytick={0.0,0.2,0.4,0.6,0.8,1.0},
            yticklabels={0.0,0.2,0.4,0.6,0.8,1.0},
            yticklabel={\pgfmathprintnumber\tick},
            yticklabel style={font=\large,
                    /pgf/number format/fixed,
            /pgf/number format/precision=1}
        },
%        point meta min=0,
%        point meta max=100,
        point meta min=0.0,
        point meta max=1.0,
        nodes near coords={\pgfmathprintnumber\pgfplotspointmeta},
        % ---------------------------------------------------------------------
        % show `nodes near coords' but adapt the style so that values
        % above a threshold get another style
        % (adapted from <http://tex.stackexchange.com/a/141006/95441>)
        % #1: the THRESHOLD after which we switch to a special display.
        nodes near coords black white/.style={
            % define the style of the nodes with "small" values
            small value/.style={
                yshift=-7pt,
%                text=white,
                text=black,
                /pgf/number format/fixed,
                /pgf/number format/precision=4,
                /pgf/number format/zerofill
            },
            % define the style of the nodes with "large" values
            large value/.style={
                yshift=-7pt,
%                text=black,
                text=white,
                /pgf/number format/fixed,
                /pgf/number format/precision=4,
                /pgf/number format/zerofill
            },
            every node near coord/.style={
                check for zero/.code={
                    \pgfmathfloatifflags{\pgfplotspointmeta}{0}{
                        % If meta=0, make the node a coordinate
                        % (which doesn't have text)
                        \pgfkeys{/tikz/coordinate}
                    }{
                        \begingroup
                        % this group is merely to switch to FPU locally.
                        % Might be unnecessary, but who knows.
                        \pgfkeys{/pgf/fpu}
                        \pgfmathparse{\pgfplotspointmeta<#1}
                        \global\let\result=\pgfmathresult
                        \endgroup
                        %
                        % simplifies debugging:
                        %\show\result
                        %
                        \pgfmathfloatcreate{1}{1.0}{0}
                        \let\ONE=\pgfmathresult
                        \ifx\result\ONE
                            % AH: our condition 'y < #1' is met.
                            \pgfkeysalso{/pgfplots/small value}
                        \else
                            % ok, proceed as usual.
                            \pgfkeysalso{/pgfplots/large value}
                        \fi
                    }
                },
                check for zero,
            },
        },
        % asign a value to the new style which is the threshold at which
        % the two style `small value' or `large value' are used
%        nodes near coords black white=50,
        nodes near coords black white=0.5,
        % -----------------------------------------------------------------
    ]
        \addplot[
            matrix plot,
            mesh/cols=20,
            point meta=explicit,draw=gray
        ] table [meta=C] {
            x y C
0 0 0.9862
1 0 0.0
2 0 0.0
3 0 0.0046
4 0 0.0
5 0 0.0
6 0 0.0
7 0 0.0
8 0 0.0046
9 0 0.0
10 0 0.0
11 0 0.0
12 0 0.0
13 0 0.0
14 0 0.0
15 0 0.0
16 0 0.0046
17 0 0.0
18 0 0.0
19 0 0.0
0 1 0.0
1 1 0.6534
2 1 0.0057
3 1 0.0114
4 1 0.0114
5 1 0.0
6 1 0.0455
7 1 0.0227
8 1 0.017
9 1 0.0
10 1 0.0852
11 1 0.0739
12 1 0.0114
13 1 0.0057
14 1 0.017
15 1 0.0057
16 1 0.0227
17 1 0.0
18 1 0.0057
19 1 0.0057
0 2 0.0
1 2 0.0114
2 2 0.7795
3 2 0.0
4 2 0.0228
5 2 0.0076
6 2 0.019
7 2 0.0076
8 2 0.0
9 2 0.0038
10 2 0.0532
11 2 0.0114
12 2 0.0038
13 2 0.0152
14 2 0.0038
15 2 0.0038
16 2 0.0456
17 2 0.0
18 2 0.0114
19 2 0.0
0 3 0.0
1 3 0.0
2 3 0.0
3 3 0.9756
4 3 0.0
5 3 0.0049
6 3 0.0098
7 3 0.0
8 3 0.0
9 3 0.0
10 3 0.0049
11 3 0.0
12 3 0.0
13 3 0.0
14 3 0.0
15 3 0.0
16 3 0.0049
17 3 0.0
18 3 0.0
19 3 0.0
0 4 0.0
1 4 0.0
2 4 0.0107
3 4 0.0
4 4 0.8556
5 4 0.0
6 4 0.0374
7 4 0.0
8 4 0.0
9 4 0.0
10 4 0.0428
11 4 0.0053
12 4 0.0107
13 4 0.0053
14 4 0.0
15 4 0.0053
16 4 0.016
17 4 0.0053
18 4 0.0
19 4 0.0053
0 5 0.0
1 5 0.0
2 5 0.0
3 5 0.0
4 5 0.0
5 5 0.9612
6 5 0.0
7 5 0.0049
8 5 0.0
9 5 0.0
10 5 0.0049
11 5 0.0
12 5 0.0049
13 5 0.0049
14 5 0.0
15 5 0.0
16 5 0.0194
17 5 0.0
18 5 0.0
19 5 0.0
0 6 0.0
1 6 0.0132
2 6 0.0088
3 6 0.0
4 6 0.0351
5 6 0.0088
6 6 0.7018
7 6 0.0088
8 6 0.0
9 6 0.0
10 6 0.0746
11 6 0.0175
12 6 0.0482
13 6 0.0
14 6 0.0263
15 6 0.0044
16 6 0.0351
17 6 0.0
18 6 0.0
19 6 0.0175
0 7 0.0
1 7 0.005
2 7 0.005
3 7 0.0
4 7 0.015
5 7 0.0
6 7 0.02
7 7 0.885
8 7 0.0
9 7 0.0
10 7 0.03
11 7 0.0
12 7 0.0
13 7 0.01
14 7 0.0
15 7 0.0
16 7 0.015
17 7 0.01
18 7 0.005
19 7 0.0
0 8 0.0
1 8 0.0
2 8 0.0
3 8 0.0
4 8 0.0036
5 8 0.0036
6 8 0.0
7 8 0.0
8 8 0.9715
9 8 0.0
10 8 0.0
11 8 0.0
12 8 0.0142
13 8 0.0
14 8 0.0036
15 8 0.0
16 8 0.0036
17 8 0.0
18 8 0.0
19 8 0.0
0 9 0.0
1 9 0.0
2 9 0.0
3 9 0.0
4 9 0.0
5 9 0.0
6 9 0.0
7 9 0.0
8 9 0.0
9 9 1.0
10 9 0.0
11 9 0.0
12 9 0.0
13 9 0.0
14 9 0.0
15 9 0.0
16 9 0.0
17 9 0.0
18 9 0.0
19 9 0.0
0 10 0.0
1 10 0.0246
2 10 0.014
3 10 0.0105
4 10 0.0316
5 10 0.0105
6 10 0.0737
7 10 0.0246
8 10 0.0
9 10 0.0
10 10 0.6035
11 10 0.0246
12 10 0.0667
13 10 0.007
14 10 0.0175
15 10 0.0035
16 10 0.0772
17 10 0.0
18 10 0.007
19 10 0.0035
0 11 0.0
1 11 0.0155
2 11 0.0116
3 11 0.0
4 11 0.0039
5 11 0.0
6 11 0.0155
7 11 0.0
8 11 0.0
9 11 0.0039
10 11 0.0271
11 11 0.8992
12 11 0.0
13 11 0.0
14 11 0.0
15 11 0.0039
16 11 0.0155
17 11 0.0
18 11 0.0039
19 11 0.0
0 12 0.0
1 12 0.0047
2 12 0.0
3 12 0.0
4 12 0.0047
5 12 0.033
6 12 0.0519
7 12 0.0047
8 12 0.0
9 12 0.0
10 12 0.0519
11 12 0.0
12 12 0.783
13 12 0.0094
14 12 0.0
15 12 0.0047
16 12 0.0425
17 12 0.0047
18 12 0.0
19 12 0.0047
0 13 0.0
1 13 0.0038
2 13 0.0227
3 13 0.0
4 13 0.0038
5 13 0.0038
6 13 0.0114
7 13 0.0038
8 13 0.0
9 13 0.0
10 13 0.0606
11 13 0.0
12 13 0.0114
13 13 0.8598
14 13 0.0
15 13 0.0
16 13 0.0189
17 13 0.0
18 13 0.0
19 13 0.0
0 14 0.0
1 14 0.0
2 14 0.0047
3 14 0.0
4 14 0.0094
5 14 0.0
6 14 0.0
7 14 0.0094
8 14 0.0
9 14 0.0
10 14 0.0188
11 14 0.0094
12 14 0.0
13 14 0.0047
14 14 0.9249
15 14 0.0047
16 14 0.0094
17 14 0.0
18 14 0.0047
19 14 0.0
0 15 0.0
1 15 0.0
2 15 0.0
3 15 0.0
4 15 0.0
5 15 0.0
6 15 0.0097
7 15 0.0
8 15 0.0
9 15 0.0
10 15 0.0097
11 15 0.0
12 15 0.0
13 15 0.0048
14 15 0.0
15 15 0.9758
16 15 0.0
17 15 0.0
18 15 0.0
19 15 0.0
0 16 0.0
1 16 0.0052
2 16 0.0078
3 16 0.0
4 16 0.0156
5 16 0.0026
6 16 0.0156
7 16 0.0104
8 16 0.0
9 16 0.0
10 16 0.0312
11 16 0.0078
12 16 0.0104
13 16 0.0052
14 16 0.0026
15 16 0.0
16 16 0.8857
17 16 0.0
18 16 0.0
19 16 0.0
0 17 0.0
1 17 0.0
2 17 0.0
3 17 0.0
4 17 0.0
5 17 0.0
6 17 0.0
7 17 0.0
8 17 0.0
9 17 0.0
10 17 0.0
11 17 0.0
12 17 0.0
13 17 0.0
14 17 0.0
15 17 0.0
16 17 0.0028
17 17 0.9972
18 17 0.0
19 17 0.0
0 18 0.0
1 18 0.0
2 18 0.0
3 18 0.0
4 18 0.0
5 18 0.0
6 18 0.0136
7 18 0.0
8 18 0.0
9 18 0.0
10 18 0.0054
11 18 0.0
12 18 0.0
13 18 0.0
14 18 0.0
15 18 0.0027
16 18 0.0
17 18 0.0
18 18 0.9783
19 18 0.0
0 19 0.0
1 19 0.0
2 19 0.0
3 19 0.0
4 19 0.0047
5 19 0.0
6 19 0.0047
7 19 0.0
8 19 0.0
9 19 0.0
10 19 0.0
11 19 0.0047
12 19 0.0
13 19 0.0
14 19 0.0
15 19 0.0
16 19 0.0
17 19 0.0
18 19 0.0
19 19 0.986

       };
    \end{axis}
\end{tikzpicture}
%
%\caption{I'm confused~5!}\label{tab:CM5}
%\end{figure*}

%% file: figures/conf_CNNMalware_Model5.tex
%\begin{tikzpicture}[scale=0.60,every node/.style={scale=0.9}]
\begin{tikzpicture}[scale=0.6]
    \begin{axis}[%colorbar/width=2.5mm,
        width=20cm,
        height=20cm,
%        colormap={blackwhite}{gray(0cm)=(1); gray(1cm)=(0.5)},
%   colormap={bluewhite}{color=(white) color=(blue)},
%   colormap={bluewhite}{color=(white) rgb255=(0,191,255)},
    colormap={bluewhite}{color=(white) rgb255=(100,149,237)},
        xticklabels={adload,agent,alureon,bho,ceeinject,cycbot,delfinject,fakerean,hotbar,lolyda,obfuscator,onlinegames,rbot,renos,startpage,vobfus,vundo,winwebsec,zbot,zeroaccess},
        xtick={0,...,19},
        xtick style={draw=none},
    xticklabel style={anchor=east,rotate=45,yshift=-5pt},
        yticklabels={adload,agent,alureon,bho,ceeinject,cycbot,delfinject,fakerean,hotbar,lolyda,obfuscator,onlinegames,rbot,renos,startpage,vobfus,vundo,winwebsec,zbot,zeroaccess},
        ytick={0,...,19},
        ytick style={draw=none},
        enlargelimits=false,
        xticklabel style={font=\large},
        yticklabel style={font=\large},
        colorbar,
        colorbar style={
%           width=0.05*\pgfkeysvalueof{/pgfplots/parent axis width},%%% added this
%           height=0.5*\pgfkeysvalueof{/pgfplots/parent axis height},
%       plot graphics/node/.style={scale=1.33,anchor=south west,inner sep=0pt,}, %%% scale colorbar fill %%%
            ytick={0.0,0.2,0.4,0.6,0.8,1.0},
            yticklabels={0.0,0.2,0.4,0.6,0.8,1.0},
            yticklabel={\pgfmathprintnumber\tick},
            yticklabel style={font=\large,
                    /pgf/number format/fixed,
            /pgf/number format/precision=1}
        },
%        point meta min=0,
%        point meta max=100,
        point meta min=0.0,
        point meta max=1.0,
        nodes near coords={\pgfmathprintnumber\pgfplotspointmeta},
        % ---------------------------------------------------------------------
        % show `nodes near coords' but adapt the style so that values
        % above a threshold get another style
        % (adapted from <http://tex.stackexchange.com/a/141006/95441>)
        % #1: the THRESHOLD after which we switch to a special display.
        nodes near coords black white/.style={
            % define the style of the nodes with "small" values
            small value/.style={
                yshift=-7pt,
%                text=white,
                text=black,
                /pgf/number format/fixed,
                /pgf/number format/precision=4,
                /pgf/number format/zerofill
            },
            % define the style of the nodes with "large" values
            large value/.style={
                yshift=-7pt,
%                text=black,
                text=white,
                /pgf/number format/fixed,
                /pgf/number format/precision=4,
                /pgf/number format/zerofill
            },
            every node near coord/.style={
                check for zero/.code={
                    \pgfmathfloatifflags{\pgfplotspointmeta}{0}{
                        % If meta=0, make the node a coordinate
                        % (which doesn't have text)
                        \pgfkeys{/tikz/coordinate}
                    }{
                        \begingroup
                        % this group is merely to switch to FPU locally.
                        % Might be unnecessary, but who knows.
                        \pgfkeys{/pgf/fpu}
                        \pgfmathparse{\pgfplotspointmeta<#1}
                        \global\let\result=\pgfmathresult
                        \endgroup
                        %
                        % simplifies debugging:
                        %\show\result
                        %
                        \pgfmathfloatcreate{1}{1.0}{0}
                        \let\ONE=\pgfmathresult
                        \ifx\result\ONE
                            % AH: our condition 'y < #1' is met.
                            \pgfkeysalso{/pgfplots/small value}
                        \else
                            % ok, proceed as usual.
                            \pgfkeysalso{/pgfplots/large value}
                        \fi
                    }
                },
                check for zero,
            },
        },
        % asign a value to the new style which is the threshold at which
        % the two style `small value' or `large value' are used
%        nodes near coords black white=50,
        nodes near coords black white=0.5,
        % -----------------------------------------------------------------
    ]
        \addplot[
            matrix plot,
            mesh/cols=20,
            point meta=explicit,draw=gray
        ] table [meta=C] {
            x y C
0 0 0.9672
1 0 0.0027
2 0 0.0027
3 0 0.0
4 0 0.0137
5 0 0.0
6 0 0.0027
7 0 0.0
8 0 0.0
9 0 0.0
10 0 0.0
11 0 0.0
12 0 0.0014
13 0 0.0
14 0 0.0014
15 0 0.0
16 0 0.0
17 0 0.0
18 0 0.0
19 0 0.0082
0 1 0.0
1 1 0.9692
2 1 0.0084
3 1 0.0
4 1 0.0112
5 1 0.0028
6 1 0.0
7 1 0.0
8 1 0.0
9 1 0.0028
10 1 0.0
11 1 0.0
12 1 0.0056
13 1 0.0
14 1 0.0
15 1 0.0
16 1 0.0
17 1 0.0
18 1 0.0
19 1 0.0
0 2 0.0028
1 2 0.0028
2 2 0.8207
3 2 0.0028
4 2 0.0812
5 2 0.0196
6 2 0.0056
7 2 0.0084
8 2 0.0
9 2 0.0056
10 2 0.0056
11 2 0.0
12 2 0.014
13 2 0.0
14 2 0.0112
15 2 0.0
16 2 0.0
17 2 0.0028
18 2 0.0
19 2 0.0168
0 3 0.0
1 3 0.0
2 3 0.0034
3 3 0.9595
4 3 0.0169
5 3 0.0068
6 3 0.0034
7 3 0.0
8 3 0.0034
9 3 0.0
10 3 0.0
11 3 0.0
12 3 0.0068
13 3 0.0
14 3 0.0
15 3 0.0
16 3 0.0
17 3 0.0
18 3 0.0
19 3 0.0
0 4 0.0262
1 4 0.0037
2 4 0.0674
3 4 0.0037
4 4 0.5955
5 4 0.0674
6 4 0.0449
7 4 0.0037
8 4 0.0037
9 4 0.0075
10 4 0.03
11 4 0.0112
12 4 0.0524
13 4 0.0
14 4 0.0075
15 4 0.0
16 4 0.0
17 4 0.0187
18 4 0.0112
19 4 0.0449
0 5 0.0038
1 5 0.0038
2 5 0.0602
3 5 0.0038
4 5 0.1015
5 5 0.7068
6 5 0.015
7 5 0.0038
8 5 0.0038
9 5 0.0
10 5 0.0038
11 5 0.0075
12 5 0.0414
13 5 0.0
14 5 0.0038
15 5 0.0
16 5 0.0
17 5 0.0075
18 5 0.015
19 5 0.0188
0 6 0.0
1 6 0.0
2 6 0.0153
3 6 0.0
4 6 0.0382
5 6 0.0
6 6 0.9084
7 6 0.0
8 6 0.0
9 6 0.0
10 6 0.0
11 6 0.0
12 6 0.0267
13 6 0.0
14 6 0.0
15 6 0.0038
16 6 0.0
17 6 0.0
18 6 0.0
19 6 0.0076
0 7 0.0039
1 7 0.0039
2 7 0.0156
3 7 0.0
4 7 0.0467
5 7 0.0117
6 7 0.0039
7 7 0.8444
8 7 0.0
9 7 0.0039
10 7 0.0117
11 7 0.0039
12 7 0.0078
13 7 0.0
14 7 0.0
15 7 0.0
16 7 0.0
17 7 0.0
18 7 0.0311
19 7 0.0117
0 8 0.0
1 8 0.0
2 8 0.0
3 8 0.0
4 8 0.0
5 8 0.0043
6 8 0.0
7 8 0.0
8 8 0.9871
9 8 0.0
10 8 0.0043
11 8 0.0
12 8 0.0
13 8 0.0043
14 8 0.0
15 8 0.0
16 8 0.0
17 8 0.0
18 8 0.0
19 8 0.0
0 9 0.0
1 9 0.0
2 9 0.0
3 9 0.0
4 9 0.0223
5 9 0.0
6 9 0.0
7 9 0.0
8 9 0.0
9 9 0.9732
10 9 0.0
11 9 0.0
12 9 0.0
13 9 0.0
14 9 0.0045
15 9 0.0
16 9 0.0
17 9 0.0
18 9 0.0
19 9 0.0
0 10 0.0091
1 10 0.0045
2 10 0.0273
3 10 0.0
4 10 0.0909
5 10 0.0182
6 10 0.0
7 10 0.0136
8 10 0.0091
9 10 0.0
10 10 0.7682
11 10 0.0
12 10 0.0182
13 10 0.0
14 10 0.0
15 10 0.0
16 10 0.0
17 10 0.0091
18 10 0.0136
19 10 0.0182
0 11 0.0
1 11 0.0
2 11 0.0184
3 11 0.0046
4 11 0.0138
5 11 0.0046
6 11 0.0
7 11 0.0
8 11 0.0046
9 11 0.0
10 11 0.0184
11 11 0.9078
12 11 0.0
13 11 0.0
14 11 0.0
15 11 0.0046
16 11 0.0
17 11 0.0046
18 11 0.0138
19 11 0.0046
0 12 0.0094
1 12 0.0
2 12 0.0704
3 12 0.0
4 12 0.1831
5 12 0.0376
6 12 0.0235
7 12 0.0
8 12 0.0047
9 12 0.0
10 12 0.0047
11 12 0.0094
12 12 0.6197
13 12 0.0
14 12 0.0
15 12 0.0
16 12 0.0
17 12 0.0
18 12 0.0141
19 12 0.0235
0 13 0.0
1 13 0.0
2 13 0.0048
3 13 0.0048
4 13 0.0
5 13 0.0048
6 13 0.0
7 13 0.0
8 13 0.0287
9 13 0.0
10 13 0.0
11 13 0.0
12 13 0.0
13 13 0.9569
14 13 0.0
15 13 0.0
16 13 0.0
17 13 0.0
18 13 0.0
19 13 0.0
0 14 0.0146
1 14 0.0097
2 14 0.0097
3 14 0.0
4 14 0.0971
5 14 0.0146
6 14 0.0388
7 14 0.0
8 14 0.0
9 14 0.0
10 14 0.0146
11 14 0.0
12 14 0.0049
13 14 0.0
14 14 0.7864
15 14 0.0
16 14 0.0
17 14 0.0
18 14 0.0
19 14 0.0097
0 15 0.0
1 15 0.0
2 15 0.0
3 15 0.0
4 15 0.0054
5 15 0.0
6 15 0.0
7 15 0.0
8 15 0.0
9 15 0.0
10 15 0.0
11 15 0.0
12 15 0.0
13 15 0.0
14 15 0.0
15 15 0.9892
16 15 0.0
17 15 0.0
18 15 0.0
19 15 0.0054
0 16 0.0
1 16 0.0164
2 16 0.0055
3 16 0.0
4 16 0.0
5 16 0.0
6 16 0.0
7 16 0.0055
8 16 0.0
9 16 0.0
10 16 0.0
11 16 0.0
12 16 0.0219
13 16 0.0
14 16 0.0
15 16 0.0
16 16 0.9344
17 16 0.0164
18 16 0.0
19 16 0.0
0 17 0.0056
1 17 0.0112
2 17 0.0056
3 17 0.0
4 17 0.0674
5 17 0.0112
6 17 0.0056
7 17 0.0
8 17 0.0
9 17 0.0056
10 17 0.0056
11 17 0.0
12 17 0.0337
13 17 0.0
14 17 0.0
15 17 0.0
16 17 0.0
17 17 0.8371
18 17 0.0
19 17 0.0112
0 18 0.0
1 18 0.0061
2 18 0.0305
3 18 0.0061
4 18 0.0671
5 18 0.0427
6 18 0.0061
7 18 0.0183
8 18 0.0244
9 18 0.0
10 18 0.0244
11 18 0.1951
12 18 0.0061
13 18 0.0
14 18 0.0
15 18 0.061
16 18 0.0
17 18 0.0
18 18 0.4939
19 18 0.0183
0 19 0.0122
1 19 0.0061
2 19 0.0671
3 19 0.0
4 19 0.3476
5 19 0.0427
6 19 0.0183
7 19 0.0061
8 19 0.0
9 19 0.0
10 19 0.0305
11 19 0.0061
12 19 0.0488
13 19 0.0
14 19 0.0061
15 19 0.0
16 19 0.0
17 19 0.0061
18 19 0.0183
19 19 0.3841
       };
    \end{axis}
\end{tikzpicture}
%
%\caption{I'm confused~5!}\label{tab:CM5}
%\end{figure*}

%% file: figures/conf_GRUMalware_Model1.tex
%\begin{tikzpicture}[scale=0.60,every node/.style={scale=0.9}]
\begin{tikzpicture}[scale=0.6]
    \begin{axis}[%colorbar/width=2.5mm,
        width=20cm,
        height=20cm,
%        colormap={blackwhite}{gray(0cm)=(1); gray(1cm)=(0.5)},
%   colormap={bluewhite}{color=(white) color=(blue)},
%   colormap={bluewhite}{color=(white) rgb255=(0,191,255)},
    colormap={bluewhite}{color=(white) rgb255=(100,149,237)},
        xticklabels={adload,agent,alureon,bho,ceeinject,cycbot,delfinject,fakerean,hotbar,lolyda,obfuscator,onlinegames,rbot,renos,startpage,vobfus,vundo,winwebsec,zbot,zeroaccess},
        xtick={0,...,19},
        xtick style={draw=none},
    xticklabel style={anchor=east,rotate=45,yshift=-5pt},
        yticklabels={adload,agent,alureon,bho,ceeinject,cycbot,delfinject,fakerean,hotbar,lolyda,obfuscator,onlinegames,rbot,renos,startpage,vobfus,vundo,winwebsec,zbot,zeroaccess},
        ytick={0,...,19},
        ytick style={draw=none},
        enlargelimits=false,
        xticklabel style={font=\large},
        yticklabel style={font=\large},
        colorbar,
        colorbar style={
%           width=0.05*\pgfkeysvalueof{/pgfplots/parent axis width},%%% added this
%           height=0.5*\pgfkeysvalueof{/pgfplots/parent axis height},
%       plot graphics/node/.style={scale=1.33,anchor=south west,inner sep=0pt,}, %%% scale colorbar fill %%%
            ytick={0.0,0.2,0.4,0.6,0.8,1.0},
            yticklabels={0.0,0.2,0.4,0.6,0.8,1.0},
            yticklabel={\pgfmathprintnumber\tick},
            yticklabel style={font=\large,
                    /pgf/number format/fixed,
            /pgf/number format/precision=1}
        },
%        point meta min=0,
%        point meta max=100,
        point meta min=0.0,
        point meta max=1.0,
        nodes near coords={\pgfmathprintnumber\pgfplotspointmeta},
        % ---------------------------------------------------------------------
        % show `nodes near coords' but adapt the style so that values
        % above a threshold get another style
        % (adapted from <http://tex.stackexchange.com/a/141006/95441>)
        % #1: the THRESHOLD after which we switch to a special display.
        nodes near coords black white/.style={
            % define the style of the nodes with "small" values
            small value/.style={
                yshift=-7pt,
%                text=white,
                text=black,
                /pgf/number format/fixed,
                /pgf/number format/precision=4,
                /pgf/number format/zerofill
            },
            % define the style of the nodes with "large" values
            large value/.style={
                yshift=-7pt,
%                text=black,
                text=white,
                /pgf/number format/fixed,
                /pgf/number format/precision=4,
                /pgf/number format/zerofill
            },
            every node near coord/.style={
                check for zero/.code={
                    \pgfmathfloatifflags{\pgfplotspointmeta}{0}{
                        % If meta=0, make the node a coordinate
                        % (which doesn't have text)
                        \pgfkeys{/tikz/coordinate}
                    }{
                        \begingroup
                        % this group is merely to switch to FPU locally.
                        % Might be unnecessary, but who knows.
                        \pgfkeys{/pgf/fpu}
                        \pgfmathparse{\pgfplotspointmeta<#1}
                        \global\let\result=\pgfmathresult
                        \endgroup
                        %
                        % simplifies debugging:
                        %\show\result
                        %
                        \pgfmathfloatcreate{1}{1.0}{0}
                        \let\ONE=\pgfmathresult
                        \ifx\result\ONE
                            % AH: our condition 'y < #1' is met.
                            \pgfkeysalso{/pgfplots/small value}
                        \else
                            % ok, proceed as usual.
                            \pgfkeysalso{/pgfplots/large value}
                        \fi
                    }
                },
                check for zero,
            },
        },
        % asign a value to the new style which is the threshold at which
        % the two style `small value' or `large value' are used
%        nodes near coords black white=50,
        nodes near coords black white=0.5,
        % -----------------------------------------------------------------
    ]
        \addplot[
            matrix plot,
            mesh/cols=20,
            point meta=explicit,draw=gray
        ] table [meta=C] {
            x y C
0 0 0.9891
1 0 0.0
2 0 0.0
3 0 0.0
4 0 0.0055
5 0 0.0
6 0 0.0
7 0 0.0
8 0 0.0
9 0 0.0
10 0 0.0014
11 0 0.0
12 0 0.0
13 0 0.0
14 0 0.0
15 0 0.0
16 0 0.0
17 0 0.0
18 0 0.0
19 0 0.0041
0 1 0.0028
1 1 0.958
2 1 0.0
3 1 0.0
4 1 0.0168
5 1 0.0
6 1 0.0
7 1 0.0
8 1 0.0
9 1 0.0056
10 1 0.0056
11 1 0.0
12 1 0.0028
13 1 0.0
14 1 0.0056
15 1 0.0
16 1 0.0
17 1 0.0
18 1 0.0
19 1 0.0028
0 2 0.0028
1 2 0.0
2 2 0.8994
3 2 0.0
4 2 0.0335
5 2 0.0056
6 2 0.0028
7 2 0.0
8 2 0.0
9 2 0.0
10 2 0.0056
11 2 0.0
12 2 0.0056
13 2 0.0
14 2 0.0
15 2 0.0
16 2 0.0
17 2 0.0112
18 2 0.0112
19 2 0.0223
0 3 0.0
1 3 0.0
2 3 0.0
3 3 0.9899
4 3 0.0068
5 3 0.0
6 3 0.0
7 3 0.0
8 3 0.0
9 3 0.0
10 3 0.0
11 3 0.0
12 3 0.0
13 3 0.0
14 3 0.0
15 3 0.0
16 3 0.0
17 3 0.0
18 3 0.0034
19 3 0.0
0 4 0.0037
1 4 0.0112
2 4 0.0337
3 4 0.0
4 4 0.7154
5 4 0.0112
6 4 0.0112
7 4 0.015
8 4 0.0
9 4 0.0037
10 4 0.0187
11 4 0.0
12 4 0.0225
13 4 0.0
14 4 0.0262
15 4 0.0
16 4 0.0
17 4 0.015
18 4 0.0225
19 4 0.0899
0 5 0.0
1 5 0.0
2 5 0.0301
3 5 0.0
4 5 0.0827
5 5 0.7782
6 5 0.015
7 5 0.0075
8 5 0.0
9 5 0.0
10 5 0.0075
11 5 0.0038
12 5 0.015
13 5 0.0
14 5 0.0113
15 5 0.0
16 5 0.0
17 5 0.0075
18 5 0.0075
19 5 0.0338
0 6 0.0
1 6 0.0
2 6 0.0152
3 6 0.0
4 6 0.0152
5 6 0.0038
6 6 0.9506
7 6 0.0
8 6 0.0
9 6 0.0
10 6 0.0
11 6 0.0
12 6 0.0
13 6 0.0
14 6 0.0038
15 6 0.0
16 6 0.0
17 6 0.0
18 6 0.0076
19 6 0.0038
0 7 0.0039
1 7 0.0
2 7 0.0
3 7 0.0039
4 7 0.0311
5 7 0.0
6 7 0.0039
7 7 0.8872
8 7 0.0
9 7 0.0
10 7 0.0078
11 7 0.0
12 7 0.0
13 7 0.0
14 7 0.0078
15 7 0.0
16 7 0.0039
17 7 0.0
18 7 0.0428
19 7 0.0078
0 8 0.0
1 8 0.0
2 8 0.0
3 8 0.0
4 8 0.0086
5 8 0.0
6 8 0.0
7 8 0.0
8 8 0.9784
9 8 0.0
10 8 0.0043
11 8 0.0
12 8 0.0043
13 8 0.0043
14 8 0.0
15 8 0.0
16 8 0.0
17 8 0.0
18 8 0.0
19 8 0.0
0 9 0.0
1 9 0.0
2 9 0.0
3 9 0.0
4 9 0.0
5 9 0.0
6 9 0.0
7 9 0.0
8 9 0.0
9 9 1.0
10 9 0.0
11 9 0.0
12 9 0.0
13 9 0.0
14 9 0.0
15 9 0.0
16 9 0.0
17 9 0.0
18 9 0.0
19 9 0.0
0 10 0.0136
1 10 0.0
2 10 0.0091
3 10 0.0
4 10 0.0909
5 10 0.0045
6 10 0.0091
7 10 0.0
8 10 0.0091
9 10 0.0
10 10 0.8
11 10 0.0091
12 10 0.0045
13 10 0.0045
14 10 0.0045
15 10 0.0
16 10 0.0
17 10 0.0
18 10 0.0045
19 10 0.0364
0 11 0.0
1 11 0.0
2 11 0.0
3 11 0.0
4 11 0.0186
5 11 0.0
6 11 0.0093
7 11 0.0
8 11 0.0
9 11 0.0
10 11 0.014
11 11 0.9349
12 11 0.0
13 11 0.0
14 11 0.0
15 11 0.0
16 11 0.0047
17 11 0.0
18 11 0.0186
19 11 0.0
0 12 0.0047
1 12 0.0
2 12 0.0
3 12 0.0
4 12 0.0751
5 12 0.0188
6 12 0.0188
7 12 0.0
8 12 0.0047
9 12 0.0047
10 12 0.0047
11 12 0.0
12 12 0.7981
13 12 0.0
14 12 0.0141
15 12 0.0
16 12 0.0
17 12 0.0094
18 12 0.0188
19 12 0.0282
0 13 0.0
1 13 0.0
2 13 0.0
3 13 0.0
4 13 0.0
5 13 0.0
6 13 0.0
7 13 0.0
8 13 0.0144
9 13 0.0
10 13 0.0
11 13 0.0048
12 13 0.0
13 13 0.9713
14 13 0.0
15 13 0.0
16 13 0.0
17 13 0.0048
18 13 0.0048
19 13 0.0
0 14 0.0
1 14 0.0
2 14 0.0
3 14 0.0
4 14 0.0194
5 14 0.0049
6 14 0.0
7 14 0.0
8 14 0.0
9 14 0.0
10 14 0.0
11 14 0.0
12 14 0.0049
13 14 0.0
14 14 0.9515
15 14 0.0
16 14 0.0
17 14 0.0
18 14 0.0146
19 14 0.0049
0 15 0.0
1 15 0.0
2 15 0.0
3 15 0.0
4 15 0.0
5 15 0.0
6 15 0.0
7 15 0.0
8 15 0.0
9 15 0.0
10 15 0.0
11 15 0.0054
12 15 0.0
13 15 0.0
14 15 0.0
15 15 0.9892
16 15 0.0
17 15 0.0
18 15 0.0054
19 15 0.0
0 16 0.0
1 16 0.0
2 16 0.0
3 16 0.0
4 16 0.0
5 16 0.0
6 16 0.0
7 16 0.0
8 16 0.0
9 16 0.0
10 16 0.0
11 16 0.0
12 16 0.0
13 16 0.0
14 16 0.0
15 16 0.0
16 16 1.0
17 16 0.0
18 16 0.0
19 16 0.0
0 17 0.0
1 17 0.0112
2 17 0.0225
3 17 0.0
4 17 0.0449
5 17 0.0
6 17 0.0
7 17 0.0
8 17 0.0
9 17 0.0
10 17 0.0
11 17 0.0
12 17 0.0056
13 17 0.0
14 17 0.0
15 17 0.0
16 17 0.0
17 17 0.8989
18 17 0.0056
19 17 0.0112
0 18 0.0
1 18 0.0
2 18 0.0244
3 18 0.0
4 18 0.0366
5 18 0.0
6 18 0.0122
7 18 0.0488
8 18 0.0
9 18 0.0
10 18 0.0183
11 18 0.2744
12 18 0.0
13 18 0.0
14 18 0.0
15 18 0.0061
16 18 0.0
17 18 0.0
18 18 0.5671
19 18 0.0122
0 19 0.0244
1 19 0.0
2 19 0.0183
3 19 0.0061
4 19 0.2195
5 19 0.0061
6 19 0.0
7 19 0.0122
8 19 0.0
9 19 0.0061
10 19 0.0244
11 19 0.0061
12 19 0.0061
13 19 0.0
14 19 0.0061
15 19 0.0
16 19 0.0
17 19 0.0122
18 19 0.0427
19 19 0.6098

       };
    \end{axis}
\end{tikzpicture}
%
%\caption{I'm confused~5!}\label{tab:CM5}
%\end{figure*}

%% file: figures/conf_StackedMalware_Model1.tex
%\begin{tikzpicture}[scale=0.60,every node/.style={scale=0.9}]
\begin{tikzpicture}[scale=0.6]
    \begin{axis}[%colorbar/width=2.5mm,
        width=20cm,
        height=20cm,
%        colormap={blackwhite}{gray(0cm)=(1); gray(1cm)=(0.5)},
%   colormap={bluewhite}{color=(white) color=(blue)},
%   colormap={bluewhite}{color=(white) rgb255=(0,191,255)},
    colormap={bluewhite}{color=(white) rgb255=(100,149,237)},
        xticklabels={adload,agent,alureon,bho,ceeinject,cycbot,delfinject,fakerean,hotbar,lolyda,obfuscator,onlinegames,rbot,renos,startpage,vobfus,vundo,winwebsec,zbot,zeroaccess},
        xtick={0,...,19},
        xtick style={draw=none},
    xticklabel style={anchor=east,rotate=45,yshift=-5pt},
        yticklabels={adload,agent,alureon,bho,ceeinject,cycbot,delfinject,fakerean,hotbar,lolyda,obfuscator,onlinegames,rbot,renos,startpage,vobfus,vundo,winwebsec,zbot,zeroaccess},
        ytick={0,...,19},
        ytick style={draw=none},
        enlargelimits=false,
        xticklabel style={font=\large},
        yticklabel style={font=\large},
        colorbar,
        colorbar style={
%           width=0.05*\pgfkeysvalueof{/pgfplots/parent axis width},%%% added this
%           height=0.5*\pgfkeysvalueof{/pgfplots/parent axis height},
%       plot graphics/node/.style={scale=1.33,anchor=south west,inner sep=0pt,}, %%% scale colorbar fill %%%
            ytick={0.0,0.2,0.4,0.6,0.8,1.0},
            yticklabels={0.0,0.2,0.4,0.6,0.8,1.0},
            yticklabel={\pgfmathprintnumber\tick},
            yticklabel style={font=\large,
                    /pgf/number format/fixed,
            /pgf/number format/precision=1}
        },
%        point meta min=0,
%        point meta max=100,
        point meta min=0.0,
        point meta max=1.0,
        nodes near coords={\pgfmathprintnumber\pgfplotspointmeta},
        % ---------------------------------------------------------------------
        % show `nodes near coords' but adapt the style so that values
        % above a threshold get another style
        % (adapted from <http://tex.stackexchange.com/a/141006/95441>)
        % #1: the THRESHOLD after which we switch to a special display.
        nodes near coords black white/.style={
            % define the style of the nodes with "small" values
            small value/.style={
                yshift=-7pt,
%                text=white,
                text=black,
                /pgf/number format/fixed,
                /pgf/number format/precision=4,
                /pgf/number format/zerofill
            },
            % define the style of the nodes with "large" values
            large value/.style={
                yshift=-7pt,
%                text=black,
                text=white,
                /pgf/number format/fixed,
                /pgf/number format/precision=4,
                /pgf/number format/zerofill
            },
            every node near coord/.style={
                check for zero/.code={
                    \pgfmathfloatifflags{\pgfplotspointmeta}{0}{
                        % If meta=0, make the node a coordinate
                        % (which doesn't have text)
                        \pgfkeys{/tikz/coordinate}
                    }{
                        \begingroup
                        % this group is merely to switch to FPU locally.
                        % Might be unnecessary, but who knows.
                        \pgfkeys{/pgf/fpu}
                        \pgfmathparse{\pgfplotspointmeta<#1}
                        \global\let\result=\pgfmathresult
                        \endgroup
                        %
                        % simplifies debugging:
                        %\show\result
                        %
                        \pgfmathfloatcreate{1}{1.0}{0}
                        \let\ONE=\pgfmathresult
                        \ifx\result\ONE
                            % AH: our condition 'y < #1' is met.
                            \pgfkeysalso{/pgfplots/small value}
                        \else
                            % ok, proceed as usual.
                            \pgfkeysalso{/pgfplots/large value}
                        \fi
                    }
                },
                check for zero,
            },
        },
        % asign a value to the new style which is the threshold at which
        % the two style `small value' or `large value' are used
%        nodes near coords black white=50,
        nodes near coords black white=0.5,
        % -----------------------------------------------------------------
    ]
        \addplot[
            matrix plot,
            mesh/cols=20,
            point meta=explicit,draw=gray
        ] table [meta=C] {
            x y C
0 0 0.9904
1 0 0.0
2 0 0.0
3 0 0.0
4 0 0.0027
5 0 0.0041
6 0 0.0
7 0 0.0
8 0 0.0
9 0 0.0
10 0 0.0014
11 0 0.0
12 0 0.0
13 0 0.0
14 0 0.0
15 0 0.0
16 0 0.0
17 0 0.0
18 0 0.0
19 0 0.0014
0 1 0.0
1 1 0.958
2 1 0.0084
3 1 0.0
4 1 0.0084
5 1 0.0028
6 1 0.0
7 1 0.0
8 1 0.0
9 1 0.0056
10 1 0.0028
11 1 0.0
12 1 0.0028
13 1 0.0
14 1 0.0084
15 1 0.0
16 1 0.0
17 1 0.0
18 1 0.0
19 1 0.0028
0 2 0.0028
1 2 0.0
2 2 0.8855
3 2 0.0056
4 2 0.0335
5 2 0.0056
6 2 0.0028
7 2 0.0028
8 2 0.0
9 2 0.0
10 2 0.0084
11 2 0.0028
12 2 0.014
13 2 0.0028
14 2 0.0
15 2 0.0
16 2 0.0
17 2 0.0056
18 2 0.0056
19 2 0.0223
0 3 0.0
1 3 0.0
2 3 0.0
3 3 0.9932
4 3 0.0034
5 3 0.0
6 3 0.0034
7 3 0.0
8 3 0.0
9 3 0.0
10 3 0.0
11 3 0.0
12 3 0.0
13 3 0.0
14 3 0.0
15 3 0.0
16 3 0.0
17 3 0.0
18 3 0.0
19 3 0.0
0 4 0.03
1 4 0.0112
2 4 0.0262
3 4 0.0
4 4 0.6779
5 4 0.03
6 4 0.0112
7 4 0.0075
8 4 0.0
9 4 0.0037
10 4 0.0225
11 4 0.0
12 4 0.0187
13 4 0.0
14 4 0.0112
15 4 0.0
16 4 0.0
17 4 0.0187
18 4 0.0262
19 4 0.1049
0 5 0.0038
1 5 0.0
2 5 0.0188
3 5 0.0038
4 5 0.0526
5 5 0.8233
6 5 0.0113
7 5 0.0038
8 5 0.0
9 5 0.0
10 5 0.0113
11 5 0.0
12 5 0.0075
13 5 0.0
14 5 0.015
15 5 0.0
16 5 0.0
17 5 0.0038
18 5 0.0113
19 5 0.0338
0 6 0.0
1 6 0.0
2 6 0.0114
3 6 0.0
4 6 0.0038
5 6 0.0038
6 6 0.9544
7 6 0.0
8 6 0.0
9 6 0.0
10 6 0.0
11 6 0.0
12 6 0.0
13 6 0.0
14 6 0.0076
15 6 0.0
16 6 0.0
17 6 0.0
18 6 0.0114
19 6 0.0076
0 7 0.0078
1 7 0.0
2 7 0.0
3 7 0.0039
4 7 0.0311
5 7 0.0
6 7 0.0039
7 7 0.8911
8 7 0.0
9 7 0.0
10 7 0.0078
11 7 0.0
12 7 0.0
13 7 0.0
14 7 0.0078
15 7 0.0
16 7 0.0039
17 7 0.0
18 7 0.035
19 7 0.0078
0 8 0.0
1 8 0.0
2 8 0.0
3 8 0.0
4 8 0.0043
5 8 0.0
6 8 0.0
7 8 0.0043
8 8 0.9784
9 8 0.0
10 8 0.0
11 8 0.0
12 8 0.0043
13 8 0.0043
14 8 0.0
15 8 0.0
16 8 0.0
17 8 0.0
18 8 0.0043
19 8 0.0
0 9 0.0
1 9 0.0
2 9 0.0
3 9 0.0
4 9 0.0045
5 9 0.0
6 9 0.0
7 9 0.0
8 9 0.0
9 9 0.9955
10 9 0.0
11 9 0.0
12 9 0.0
13 9 0.0
14 9 0.0
15 9 0.0
16 9 0.0
17 9 0.0
18 9 0.0
19 9 0.0
0 10 0.0045
1 10 0.0
2 10 0.0227
3 10 0.0
4 10 0.0682
5 10 0.0
6 10 0.0091
7 10 0.0091
8 10 0.0045
9 10 0.0
10 10 0.8045
11 10 0.0045
12 10 0.0045
13 10 0.0
14 10 0.0
15 10 0.0
16 10 0.0
17 10 0.0
18 10 0.0227
19 10 0.0455
0 11 0.0
1 11 0.0
2 11 0.0047
3 11 0.0
4 11 0.0093
5 11 0.0
6 11 0.0093
7 11 0.0047
8 11 0.0
9 11 0.0
10 11 0.0
11 11 0.9302
12 11 0.0047
13 11 0.0
14 11 0.0047
15 11 0.0
16 11 0.0
17 11 0.0
18 11 0.0233
19 11 0.0093
0 12 0.0047
1 12 0.0
2 12 0.0282
3 12 0.0
4 12 0.0282
5 12 0.0516
6 12 0.0141
7 12 0.0
8 12 0.0
9 12 0.0
10 12 0.0047
11 12 0.0
12 12 0.831
13 12 0.0
14 12 0.0094
15 12 0.0
16 12 0.0
17 12 0.0
18 12 0.0047
19 12 0.0235
0 13 0.0
1 13 0.0
2 13 0.0
3 13 0.0
4 13 0.0
5 13 0.0
6 13 0.0
7 13 0.0
8 13 0.0
9 13 0.0
10 13 0.0048
11 13 0.0048
12 13 0.0048
13 13 0.9856
14 13 0.0
15 13 0.0
16 13 0.0
17 13 0.0
18 13 0.0
19 13 0.0
0 14 0.0049
1 14 0.0
2 14 0.0194
3 14 0.0
4 14 0.0243
5 14 0.0146
6 14 0.0
7 14 0.0
8 14 0.0
9 14 0.0
10 14 0.0
11 14 0.0049
12 14 0.0
13 14 0.0
14 14 0.9126
15 14 0.0049
16 14 0.0
17 14 0.0
18 14 0.0049
19 14 0.0097
0 15 0.0
1 15 0.0
2 15 0.0
3 15 0.0
4 15 0.0
5 15 0.0
6 15 0.0
7 15 0.0
8 15 0.0
9 15 0.0
10 15 0.0054
11 15 0.0
12 15 0.0
13 15 0.0
14 15 0.0
15 15 0.9892
16 15 0.0
17 15 0.0
18 15 0.0054
19 15 0.0
0 16 0.0
1 16 0.0
2 16 0.0
3 16 0.0
4 16 0.0
5 16 0.0
6 16 0.0
7 16 0.0
8 16 0.0
9 16 0.0
10 16 0.0
11 16 0.0
12 16 0.0
13 16 0.0
14 16 0.0
15 16 0.0
16 16 1.0
17 16 0.0
18 16 0.0
19 16 0.0
0 17 0.0
1 17 0.0
2 17 0.0112
3 17 0.0
4 17 0.0393
5 17 0.0
6 17 0.0
7 17 0.0112
8 17 0.0
9 17 0.0
10 17 0.0
11 17 0.0056
12 17 0.0056
13 17 0.0
14 17 0.0056
15 17 0.0
16 17 0.0
17 17 0.9045
18 17 0.0
19 17 0.0169
0 18 0.0
1 18 0.0061
2 18 0.0122
3 18 0.0
4 18 0.0488
5 18 0.0
6 18 0.0122
7 18 0.0488
8 18 0.0
9 18 0.0
10 18 0.0183
11 18 0.2744
12 18 0.0
13 18 0.0
14 18 0.0
15 18 0.0061
16 18 0.0
17 18 0.0
18 18 0.561
19 18 0.0122
0 19 0.0671
1 19 0.0
2 19 0.0427
3 19 0.0061
4 19 0.1646
5 19 0.0366
6 19 0.0
7 19 0.0305
8 19 0.0
9 19 0.0
10 19 0.0427
11 19 0.0122
12 19 0.0061
13 19 0.0
14 19 0.0
15 19 0.0
16 19 0.0
17 19 0.0061
18 19 0.0366
19 19 0.5488

       };
    \end{axis}
\end{tikzpicture}
%
%\caption{I'm confused~5!}\label{tab:CM5}
%\end{figure*}

%% file: figures/conf_tl_vgg19.tex
%\begin{tikzpicture}[scale=0.60,every node/.style={scale=0.9}]
\begin{tikzpicture}[scale=0.6]
    \begin{axis}[%colorbar/width=2.5mm,
        width=20cm,
        height=20cm,
%        colormap={blackwhite}{gray(0cm)=(1); gray(1cm)=(0.5)},
%   colormap={bluewhite}{color=(white) color=(blue)},
%   colormap={bluewhite}{color=(white) rgb255=(0,191,255)},
    colormap={bluewhite}{color=(white) rgb255=(100,149,237)},
        xticklabels={adload,agent,alureon,bho,ceeinject,cycbot,delfinject,fakerean,hotbar,lolyda,obfuscator,onlinegames,rbot,renos,startpage,vobfus,vundo,winwebsec,zbot,zeroaccess},
        xtick={0,...,19},
        xtick style={draw=none},
    xticklabel style={anchor=east,rotate=45,yshift=-5pt},
        yticklabels={adload,agent,alureon,bho,ceeinject,cycbot,delfinject,fakerean,hotbar,lolyda,obfuscator,onlinegames,rbot,renos,startpage,vobfus,vundo,winwebsec,zbot,zeroaccess},
        ytick={0,...,19},
        ytick style={draw=none},
        enlargelimits=false,
        xticklabel style={font=\large},
        yticklabel style={font=\large},
        colorbar,
        colorbar style={
%           width=0.05*\pgfkeysvalueof{/pgfplots/parent axis width},%%% added this
%           height=0.5*\pgfkeysvalueof{/pgfplots/parent axis height},
%       plot graphics/node/.style={scale=1.33,anchor=south west,inner sep=0pt,}, %%% scale colorbar fill %%%
            ytick={0.0,0.2,0.4,0.6,0.8,1.0},
            yticklabels={0.0,0.2,0.4,0.6,0.8,1.0},
            yticklabel={\pgfmathprintnumber\tick},
            yticklabel style={font=\large,
                    /pgf/number format/fixed,
            /pgf/number format/precision=1}
        },
%        point meta min=0,
%        point meta max=100,
        point meta min=0.0,
        point meta max=1.0,
        nodes near coords={\pgfmathprintnumber\pgfplotspointmeta},
        % ---------------------------------------------------------------------
        % show `nodes near coords' but adapt the style so that values
        % above a threshold get another style
        % (adapted from <http://tex.stackexchange.com/a/141006/95441>)
        % #1: the THRESHOLD after which we switch to a special display.
        nodes near coords black white/.style={
            % define the style of the nodes with "small" values
            small value/.style={
                yshift=-7pt,
%                text=white,
                text=black,
                /pgf/number format/fixed,
                /pgf/number format/precision=4,
                /pgf/number format/zerofill
            },
            % define the style of the nodes with "large" values
            large value/.style={
                yshift=-7pt,
%                text=black,
                text=white,
                /pgf/number format/fixed,
                /pgf/number format/precision=4,
                /pgf/number format/zerofill
            },
            every node near coord/.style={
                check for zero/.code={
                    \pgfmathfloatifflags{\pgfplotspointmeta}{0}{
                        % If meta=0, make the node a coordinate
                        % (which doesn't have text)
                        \pgfkeys{/tikz/coordinate}
                    }{
                        \begingroup
                        % this group is merely to switch to FPU locally.
                        % Might be unnecessary, but who knows.
                        \pgfkeys{/pgf/fpu}
                        \pgfmathparse{\pgfplotspointmeta<#1}
                        \global\let\result=\pgfmathresult
                        \endgroup
                        %
                        % simplifies debugging:
                        %\show\result
                        %
                        \pgfmathfloatcreate{1}{1.0}{0}
                        \let\ONE=\pgfmathresult
                        \ifx\result\ONE
                            % AH: our condition 'y < #1' is met.
                            \pgfkeysalso{/pgfplots/small value}
                        \else
                            % ok, proceed as usual.
                            \pgfkeysalso{/pgfplots/large value}
                        \fi
                    }
                },
                check for zero,
            },
        },
        % asign a value to the new style which is the threshold at which
        % the two style `small value' or `large value' are used
%        nodes near coords black white=50,
        nodes near coords black white=0.5,
        % -----------------------------------------------------------------
    ]
        \addplot[
            matrix plot,
            mesh/cols=20,
            point meta=explicit,draw=gray
        ] table [meta=C] {
            x y C
0 0 0.9852
1 0 0.0
2 0 0.0
3 0 0.0049
4 0 0.0
5 0 0.0
6 0 0.0
7 0 0.0
8 0 0.0
9 0 0.0
10 0 0.0
11 0 0.0
12 0 0.0
13 0 0.0
14 0 0.0099
15 0 0.0
16 0 0.0
17 0 0.0
18 0 0.0
19 0 0.0
0 1 0.0
1 1 0.8261
2 1 0.0186
3 1 0.0
4 1 0.0
5 1 0.0
6 1 0.0248
7 1 0.0062
8 1 0.0
9 1 0.0
10 1 0.0186
11 1 0.0248
12 1 0.0062
13 1 0.0
14 1 0.0248
15 1 0.0062
16 1 0.0186
17 1 0.0124
18 1 0.0062
19 1 0.0062
0 2 0.0
1 2 0.0
2 2 0.9308
3 2 0.0
4 2 0.0115
5 2 0.0
6 2 0.0038
7 2 0.0
8 2 0.0
9 2 0.0038
10 2 0.0154
11 2 0.0038
12 2 0.0038
13 2 0.0
14 2 0.0
15 2 0.0
16 2 0.0231
17 2 0.0
18 2 0.0038
19 2 0.0
0 3 0.0
1 3 0.0
2 3 0.0
3 3 0.9691
4 3 0.0
5 3 0.0
6 3 0.0
7 3 0.0
8 3 0.0
9 3 0.0
10 3 0.0116
11 3 0.0039
12 3 0.0039
13 3 0.0039
14 3 0.0039
15 3 0.0
16 3 0.0
17 3 0.0039
18 3 0.0
19 3 0.0
0 4 0.0
1 4 0.0055
2 4 0.0331
3 4 0.0
4 4 0.8508
5 4 0.0
6 4 0.0387
7 4 0.0
8 4 0.0
9 4 0.0
10 4 0.0221
11 4 0.0055
12 4 0.0055
13 4 0.011
14 4 0.0
15 4 0.0
16 4 0.0276
17 4 0.0
18 4 0.0
19 4 0.0
0 5 0.0
1 5 0.0
2 5 0.0
3 5 0.0
4 5 0.0
5 5 0.9893
6 5 0.0
7 5 0.0
8 5 0.0
9 5 0.0
10 5 0.0
11 5 0.0
12 5 0.0
13 5 0.0
14 5 0.0
15 5 0.0
16 5 0.0107
17 5 0.0
18 5 0.0
19 5 0.0
0 6 0.0
1 6 0.0
2 6 0.0082
3 6 0.0041
4 6 0.0247
5 6 0.0
6 6 0.786
7 6 0.0082
8 6 0.0082
9 6 0.0
10 6 0.0576
11 6 0.0
12 6 0.0165
13 6 0.0247
14 6 0.0082
15 6 0.0
16 6 0.0412
17 6 0.0041
18 6 0.0082
19 6 0.0
0 7 0.0
1 7 0.0045
2 7 0.009
3 7 0.0
4 7 0.009
5 7 0.0045
6 7 0.009
7 7 0.8643
8 7 0.0
9 7 0.0
10 7 0.0271
11 7 0.0
12 7 0.0045
13 7 0.0136
14 7 0.009
15 7 0.0
16 7 0.0362
17 7 0.0045
18 7 0.0
19 7 0.0045
0 8 0.0
1 8 0.0034
2 8 0.0034
3 8 0.0
4 8 0.0
5 8 0.0
6 8 0.0
7 8 0.0
8 8 0.9897
9 8 0.0
10 8 0.0
11 8 0.0
12 8 0.0034
13 8 0.0
14 8 0.0
15 8 0.0
16 8 0.0
17 8 0.0
18 8 0.0
19 8 0.0
0 9 0.0
1 9 0.0
2 9 0.0
3 9 0.0
4 9 0.0
5 9 0.0
6 9 0.0
7 9 0.0
8 9 0.0
9 9 1.0
10 9 0.0
11 9 0.0
12 9 0.0
13 9 0.0
14 9 0.0
15 9 0.0
16 9 0.0
17 9 0.0
18 9 0.0
19 9 0.0
0 10 0.0
1 10 0.0212
2 10 0.0459
3 10 0.0035
4 10 0.0283
5 10 0.0035
6 10 0.053
7 10 0.0071
8 10 0.0
9 10 0.0
10 10 0.6466
11 10 0.0035
12 10 0.0389
13 10 0.0389
14 10 0.0106
15 10 0.0035
16 10 0.0742
17 10 0.0106
18 10 0.0035
19 10 0.0071
0 11 0.0
1 11 0.0102
2 11 0.0102
3 11 0.0
4 11 0.0034
5 11 0.0
6 11 0.0102
7 11 0.0
8 11 0.0
9 11 0.0
10 11 0.0068
11 11 0.9288
12 11 0.0034
13 11 0.0136
14 11 0.0034
15 11 0.0
16 11 0.0068
17 11 0.0
18 11 0.0034
19 11 0.0
0 12 0.0046
1 12 0.0
2 12 0.0092
3 12 0.0046
4 12 0.0276
5 12 0.0046
6 12 0.0323
7 12 0.0
8 12 0.0046
9 12 0.0
10 12 0.0507
11 12 0.0092
12 12 0.7696
13 12 0.0276
14 12 0.0092
15 12 0.0
16 12 0.0415
17 12 0.0046
18 12 0.0
19 12 0.0
0 13 0.0
1 13 0.0
2 13 0.0
3 13 0.0
4 13 0.0
5 13 0.0
6 13 0.0038
7 13 0.0038
8 13 0.0
9 13 0.0
10 13 0.0
11 13 0.0
12 13 0.0115
13 13 0.9615
14 13 0.0038
15 13 0.0038
16 13 0.0115
17 13 0.0
18 13 0.0
19 13 0.0
0 14 0.0
1 14 0.0129
2 14 0.0
3 14 0.0
4 14 0.0086
5 14 0.0
6 14 0.0
7 14 0.0
8 14 0.0
9 14 0.0
10 14 0.0086
11 14 0.0043
12 14 0.0
13 14 0.0043
14 14 0.9571
15 14 0.0
16 14 0.0043
17 14 0.0
18 14 0.0
19 14 0.0
0 15 0.0
1 15 0.0057
2 15 0.0
3 15 0.0
4 15 0.0
5 15 0.0
6 15 0.0057
7 15 0.0
8 15 0.0
9 15 0.0
10 15 0.0
11 15 0.0057
12 15 0.0115
13 15 0.0
14 15 0.0
15 15 0.9598
16 15 0.0115
17 15 0.0
18 15 0.0
19 15 0.0
0 16 0.0
1 16 0.0
2 16 0.0085
3 16 0.0
4 16 0.0057
5 16 0.0028
6 16 0.0085
7 16 0.0028
8 16 0.0
9 16 0.0
10 16 0.0255
11 16 0.0057
12 16 0.0085
13 16 0.0113
14 16 0.0057
15 16 0.0028
16 16 0.9065
17 16 0.0028
18 16 0.0
19 16 0.0028
0 17 0.0
1 17 0.0
2 17 0.0
3 17 0.0
4 17 0.0
5 17 0.0
6 17 0.0
7 17 0.0
8 17 0.0
9 17 0.0
10 17 0.0
11 17 0.0
12 17 0.0
13 17 0.0
14 17 0.0
15 17 0.0
16 17 0.0
17 17 1.0
18 17 0.0
19 17 0.0
0 18 0.0
1 18 0.0
2 18 0.0028
3 18 0.0
4 18 0.0
5 18 0.0
6 18 0.0028
7 18 0.0
8 18 0.0
9 18 0.0
10 18 0.0057
11 18 0.0
12 18 0.0
13 18 0.0028
14 18 0.0
15 18 0.0
16 18 0.0028
17 18 0.0
18 18 0.983
19 18 0.0
0 19 0.0
1 19 0.0
2 19 0.0
3 19 0.0
4 19 0.0
5 19 0.0
6 19 0.0
7 19 0.0
8 19 0.0
9 19 0.0
10 19 0.0
11 19 0.0049
12 19 0.0
13 19 0.0049
14 19 0.0
15 19 0.0049
16 19 0.0
17 19 0.0
18 19 0.0
19 19 0.9854

       };
    \end{axis}
\end{tikzpicture}
%
%\caption{I'm confused~5!}\label{tab:CM5}
%\end{figure*}